%% file: fintopos3.tex
\documentclass[12pt,twoside]{article}
\usepackage{amsmath, amsfonts, amscd, amssymb}
\usepackage{amsthm}
\usepackage{fancybox, eufrak, euscript, oldgerm}
\newcommand{\aconn}{\mathcal{A}}

\newcommand{\aut}{{\mathcal{A}}ut}

\newcommand{\base}{\mathfrak{B}}
\newcommand{\basis}{\mathcal{B}}
\newcommand{\cat}{\mathfrak{C}}

\newcommand{\conn}{\mathcal{D}}

\newcommand{\cov}{\mathfrak{U}}
\newcommand{\curv}{R}
\newcommand{\ric}{{\mathcal{R}}}

\newcommand{\ricci}{\EuScript{R}}
\newcommand{\corv}{\mathfrak{R}}

\newcommand{\eh}{\mathfrak{E}\mathfrak{H}}

\newcommand{\gauge}{\mathcal{U}}

\newcommand{\morph}{\EuScript{F}}
\newcommand{\geomorph}{\EuScript{G}\EuScript{M}}
\newcommand{\hil}{H}
\newcommand{\hilb}{\mathcal{H}}
\newcommand{\hilbert}{\mathfrak{H}_{fcq}}
\newcommand{\Hom}{{\mathcal{H}}om}
\newcommand{\kd}{\text{\texttt{d}}}

\newcommand{\inv}{\overleftarrow{\EuScript{P}}}
\newcommand{\invf}{\overleftarrow{\cov}_{i}}
\newcommand{\invomg}{\overleftarrow{\Omega}}
\newcommand{\inveinst}{\overleftarrow{\EuScript{E}}}

\newcommand{\diromg}{\overrightarrow{\Omega}}

\newcommand{\lsh}{{\mathcal{L}}}
\newcommand{\loc}{\lsh oc}

\newcommand{\man}{{\mathcal{M}}an}
\newcommand{\modl}{\mathbf{\mathcal{E}}}

\newcommand{\natf}{\EuScript{N}}

\newcommand{\omg}{\Omega}
\newcommand{\Omg}{\mathbf{\Omega}}

\newcommand{\poset}{{\mathcal{P}}{\mathcal{O}}}

\newcommand{\cont}{\mathcal{C}^{0}}
\newcommand{\smooth}{\mathcal{C}^{\infty}}
\newcommand{\sconn}{\textsf{A}}

\newcommand{\set}{{\mathbf{Set}}}
\newcommand{\sh}{{\mathcal{S}}\mathbf{hv}}

\newcommand{\struc}{\mathbf{A}}

\newcommand{\sub}{\Omega_{fcq}}
\newcommand{\triad}{{\mathfrak{T}}}
\newcommand{\ctriad}{{\mathfrak{D}}\triad}
\newcommand{\fintriad}{\ctriad_{fcq}}

\newcommand{\bull}{{\scriptstyle\bullet}}

\newcommand{\com}{\mathbb{C}}
\newcommand{\mapto}{\longrightarrow}

\newcommand{\Z}{\mathbb{Z}}
\newcommand{\R}{\mathbb{R}}
\newcommand{\K}{\mathbb{K}}

\input{diagram}

\title{\bf Finitary Topos for Locally Finite, Causal\\and Quantal Vacuum
Einstein Gravity\thanks{Posted at the {\em General Relativity and
Quantum Cosmology} (gr-qc) electronic archive ({\bf
www.arXiv.org}), as: gr-qc/0507100.}}

\author{Ioannis Raptis\thanks{European Commission Marie Curie Reintegration Research Fellow, Algebra and Geometry Section,
Department of Mathematics, University of Athens,
Panepistimioupolis, Athens 157 84, Greece; \underline{\em and}
Visiting Researcher, Theoretical Physics Group, Blackett
Laboratory, Imperial College of Science, Technology and Medicine,
Prince Consort Road, South Kensington, London SW7 2BZ, UK; e-mail:
i.raptis@ic.ac.uk}}

\date{\today}

\begin{document}

{\catcode`\ =13\global\let =\ \catcode`\^^M=13
\gdef^^M{\par\noindent}}
\def\verbatim{\tt
\catcode`\^^M=13 \catcode`\ =13 \catcode`\\=12 \catcode`\{=12
\catcode`\}=12 \catcode`\_=12 \catcode`\^=12 \catcode`\&=12
\catcode`\~=12 \catcode`\#=12 \catcode`\%=12 \catcode`\$=12
\catcode`|=0 }

\maketitle

\pagestyle{myheadings}\markboth{\centerline {\small {\sc {Ioannis
Raptis}}}}{\centerline {\footnotesize {\sc {Finitary Topos
Structure for Locally Finite, Causal and Quantal Vacuum Einstein
Gravity}}}}

\pagenumbering{arabic}

\begin{abstract}

\noindent The pentalogy
\cite{malrap1,malrap2,malrap3,rap5,malrap4} is brought to its
categorical climax by organizing the curved finitary spacetime
sheaves of quantum causal sets involved therein, on which a
finitary (:locally finite), singularity-free, background manifold
independent and geometrically prequantized version of the
gravitational vacuum Einstein field equations were seen to hold,
into a topos structure $\fintriad$. We show that the category of
finitary differential triads $\fintriad$ is a finitary instance of
an elementary topos proper in the original sense due to Lawvere
and Tierney. We present in the light of Abstract Differential
Geometry (ADG) a Grothendieck-type of generalization of Sorkin's
finitary substitutes of continuous spacetime manifold topologies,
the latter's topological refinement inverse systems of locally
finite coverings and their associated coarse graining sieves, the
upshot being that $\fintriad$ is also a finitary example of a
Grothendieck topos. In the process, we discover that the subobject
classifier $\sub$ of $\fintriad$ is a Heyting algebra type of
object, thus we infer that the internal logic of our finitary
topos is intuitionistic, as expected. We also introduce the new
notion of `finitary differential geometric morphism' which, as
befits ADG, gives a differential geometric slant to Sorkin's
purely topological acts of refinement (:coarse graining). Based on
finitary differential geometric morphisms regarded as natural
transformations of the relevant sheaf categories, we observe that
the functorial ADG-theoretic version of the principle of general
covariance of General Relativity is preserved under topological
refinement. The paper closes with a thorough discussion of four
future routes we could take in order to further develop our
topos-theoretic perspective on ADG-gravity along certain
categorical trends in current quantum gravity research.

\vskip 0.2in



\end{abstract}

\newpage

\tableofcontents

\newpage

\setlength{\textwidth}{15.9cm} 
\setlength{\oddsidemargin}{0cm}  
\setlength{\evensidemargin}{0cm} 
\setlength{\topskip}{0pt}  
\setlength{\textheight}{21.8cm} 
\setlength{\footskip}{-2cm}

\setlength{\topmargin}{0pt}

\section{Prologue cum Physical Motivation: the Past and the Present}

In the past decade or so, we have witnessed vigorous activity in
various applications of categorical---in particular, (pre)sheaf
and topos-theoretic \cite{macmo}---ideas to Quantum Theory (QT)
and Quantum Gravity (QG).

With respect to QT proper, topos theory appears to be a suitable
and elegant framework in which to express the non-objective,
non-classical ({\it ie}, non-Boolean), so-called `neo-realist'
({\it ie}, intuitionistic), and contextual underpinnings of the
logic of (non-relativistic) Quantum Mechanics (QM), as manifested
for example by the Kochen-Specker theorem in standard quantum
logic \cite{buttish1,buttish2,buttish3,buttish4,rawling}.
Recently, Isham {\it et al.}'s topos perspective on the
Kochen-Specker theorem and the Boolean algebra-localized
(:contextualized) logic of QT has triggered research on applying
category-theoretic ideas to the `problem' of non-trivial
localization properties of quantum observables \cite{zafiris0}.
Topos theory has also been used to reveal the intuitionistic
colors of the logic underlying the `non-instrumentalist',
non-Copenhagean, `quantum state collapse-free' consistent
histories approach to QM \cite{ish4}.

At the same time, topos theory has also been applied to General
Relativity (GR), especially by the Siberian school of
`toposophers' \cite{guts0,guts1,guts2,guts3,guts,grink}. Emphasis
here is placed on using the intuitionistic-type of internal logic
of a so-called `{\em formal smooth topos}', which is assumed to
replace the (category of finite-dimensional) smooth spacetime
manifold(s) of GR, in order to define a new kind of differential
geometry more general than the Classical ({\it ie}, from a logical
standpoint, Boolean topos $\set$-based) Differential Geometry
(CDG) of finite-dimensional differential (:$\smooth$-smooth)
manifolds. The tacit assumption here is that the standard
kinematical structure of GR---the background pseudo-Riemannian
smooth spacetime manifold---is basically ({\it ie}, when stripped
of its topological, differential and smooth Lorentzian metric
structures) a classical point-set continuum living in the topos
$\set$ of `constant' sets, with its `innate' Boolean (:classical)
logic \cite{gold,macmo}. This new `intuitionistic differential
calculus' pertains to the celebrated Synthetic Differential
Geometry (SDG) of Kock and Lawvere \cite{kock,laven}, in terms of
which the differential equations of gravity (:Einstein equations)
are then formulated in a `formal smooth manifold'. A byproduct of
this perspective on gravity is that the causal structure (:`causal
topology') of the pointed spacetime continuum of GR is also
revised, being replaced by an axiomatic scheme of `pointless
regions' and coverings for them recalling Grothendieck's
pioneering work on generalized topological spaces called {\em
sites} and their associated sheaf categories (:topoi), which
culminated in the study of new, abstract (sheaf) cohomology
theories in modern algebraic geometry \cite{macmo}.

Arguably however, the ultimate challenge for theoretical physics
research in the new millennium is to arrive at a conceptually
sound and calculationally sensible ({\it ie}, finite) QG---the
traditionally supposed (and expected!) marriage of QT with GR.
Here too, category, (pre)sheaf and topos theory has been
anticipated to play a central role for many different reasons, due
to various different motivations, and with different aims in mind,
depending on the approach to QG that one favors
\cite{crane0,trifonov,buttish4,marko,ish3,rap4,ish5,ish6,ish7,ish8,kato,kato1,rap3,rap4,rap6}.

Akin to the present work is the recent paper of Christensen and
Crane on so-called `{\em causal sites}' (causites) \cite{crane}.
Like the Novosibirsk endeavors in classical GR mentioned above,
this is an axiomatic looking scheme based on Grothendieck-type of
$2$-categories (:$2$-sites) in which the topological and causal
structure of spacetime are intimately entwined and, when endowed
with some suitable finiteness conditions, appear to be well
prepared for quantization using combinatory-topological state-sum
models coming from Relativistic Spin Networks and Topological
Quantum Field Theory \cite{barcrane}. Ultimately, the theory
aspires to lead to a finite theory of quantum spacetime geometry
and QGR in a point-free $2$-topos theoretic setting.

In the present paper too, we extend by topos-theoretic means
previous work on applying Mallios' purely algebraic
(:sheaf-theoretic) and background differential manifold
independent Abstract Differential Geometry (ADG)
\cite{mall1,mall2,mall4} towards formulating a finitary (:locally
finite), causal and quantal, as well as singularities-{\it
cum}-infinities free, version of Lorentzian (vacuum) Einstein
gravity \cite{malrap1,malrap2,malrap3,rap5,malrap4}. This
extension is accomplished by organizing the curved finitary
spacetime sheaves (finsheaves) of quantum causal sets (qausets)
involved therein, on which a finitary, singularities and
infinities-free, background manifold independent and geometrically
(pre)quantized version of the gravitational (vacuum) Einstein
field equations were seen to hold, into a `{\em finitary topos}'
(fintopos) structure $\fintriad$.\footnote{As in the previous
pentalogy \cite{malrap1,malrap2,malrap3,rap5,malrap4}, the
subscript `$fcq$' is an acronym standing for ($f$)initary,
($c$)ausal and ($q$)uantal. Occasionally we shall augment this
3-letter acronym with a fourth letter, `$v$', standing for
($v$)acuum. The general ADG-theoretic perspective on (vacuum
Einstein) gravity may be coined `{\em ADG-gravity}' for short
\cite{rap5,malrap4}. {\it In toto}, the theory propounded in
\cite{malrap3,rap5,malrap4}, and topos-theoretically extended
herein, may be called `$fcqv$-ADG-gravity'.}

The key observation supporting this topos organization of the said
finsheaves is that the category $\fintriad$ of finitary
differential triads (fintriads)---{\em the} basic structural units
on which our application of ADG to the finitary spacetime regime
rests---is a finitary instance of an {\em elementary topos} (ET)
in the original sense due to Lawvere and Tierney \cite{macmo}.
This result is a straightforward one coming from recent thorough
investigations of Papatriantafillou about the general categorical
properties of the (abstract) category of differential triads
$\ctriad$ \cite{pap1,pap2,pap3,pap4,pap5}, of which $\fintriad$ is
a concrete and full subcategory.

There is also another way of showing that $\fintriad$ is a topos.
From the finitary stance that we have adopted throughout our
applications of ADG-theoretic ideas to spacetime and gravity
\cite{malrap1,malrap2,malrap3,rap5,malrap4}, we will show that
$\fintriad$ is a finitary example of a {\em Grothendieck topos}
(GT) \cite{macmo}. This arises from a general, Grothendieck-type
of perspective on the finitary (open) coverings and their
associated locally finite partially ordered set (poset)
substitutes of continuous (spacetime) manifolds originally due to
Sorkin \cite{sork0}. The main structures involved here are what
one might call `{\em covering coarse graining sieves}' adapted to
the said finitary open covers (fincovers) and their associated
locally finite posets. These finitary sieves (finsieves) are
easily seen to define (:generate) a {\em Grothendieck topology} on
the poset of all open subsets of the topological spacetime
manifold $X$, which, in turn, when regarded as a poset category,
is turned into a {\em site}---{\it ie}, a category endowed with a
{\em Grothendieck topology} \cite{macmo}. Then, the well known
result of topos theory is evoked, namely, that the collection
$\fintriad$ of all the said finsheaves over this site is a
finitary instance of a GT. Of course, it is a general result that
every GT is an ET \cite{macmo}, thus $\fintriad$ qualifies as
both.\footnote{For similar Grothendieck-type of ideas in an
ADG-theoretic setting, but with different physico-mathematical
motivations and aims, the reader is referred to a recent paper by
Zafiris \cite{zafiris1}, which builds on the aforementioned work
on algebraic quantum observables' localizations \cite{zafiris0}.}

Much in the same way that the locally finite posets in
\cite{sork0} were regarded as finitary substitutes of the
continuous topology of the topological spacetime manifold $X$ and,
similarly, the finsheaves in \cite{rap2} as finitary replacements
of the sheaf $\cont_{X}$ of continuous functions on $X$,
$\fintriad$ may be viewed as a finitary approximation of the
elementary-Grothendieck topos (EGT) $\sh^{0}(X)$---the category of
sheaves of (rings of) continuous functions over the base
$\cont$-manifold $X$ \cite{macmo}. Moreover, since the
construction of our fintopos employs the basic ADG concepts and
technology, $\fintriad$ has not only topological, but also {\em
differential geometric} attributes and significance, and thus it
may be thought of as a finitary substitute of the category $\man$
of finite-dimensional differential manifolds---a category that
{\em cannot} be viewed as a topos proper. As we shall argue in the
present paper, this is just one instance of the categorical
versatility and import of ADG.

Furthermore, in a technical sense, since the EG fintopos
$\fintriad$ is manifestly ({\it ie}, by construction) {\em
finitely generated}, it is both {\em coherent} and {\em localic}
\cite{macmo}. The underlying locale is the usual lattice of open
subsets of the pointed, base topological manifold $X$ that Sorkin
initially considered in \cite{sork0}. This gives us important
clues about what is the {\em subobject classifier} \cite{macmo}
$\sub$ of $\fintriad$. Also, being coherent, $\fintriad$ has
enough points \cite{macmo}. Indeed, these are the points (of $X$)
that Sorkin initially `blew up' or `smeared out' by open subsets
about them, being physically motivated by the observation that a
point is an (operationally) `ideal' entity with pathological
(:`singular') behavior in GR. Parenthetically, and from a physical
viewpoint, the ideal ({\it ie}, non-pragmatic) character of
spacetime points is reflected by the apparent theoretical
impossibility to localize physical fields over them. Indeed, as
also noted in \cite{sork1}, a conspiracy between the equivalence
principle of GR and the uncertainty principle of QM appears to
prohibit the infinite point-localization of the gravitational
field in the sense that the more one tries to localize (:measure)
the gravitational field, the more (microscopic)
energy-mass-momentum probes one is forced to use, which in turn
produce a gravitational field strong enough to perturb
uncontrollably and without bound the original field that one
initially set out to measure. In geometrical space-time imagery,
one cannot localize the gravitational field more sharply than a
so-called Planck length-time (in which both the quantum of action
$\hbar$ and Newton's gravitational constant $G$ are involved)
without creating a black hole, which fuzzies or blurs out things
so to speak. Thus, Sorkin substituted points by `regions' (:open
sets) about them, hence also, effectively, the pointed $X$---with
the usual Euclidean $\cont$-topology ``{\em carried by its
points}'' \cite{sork0}---was replaced by the `pointless locale'
\cite{macmo} of its open subsets. Of course, Sorkin also provided
a mechanism---technically, a projective limit procedure---for
recovering (the ideal points of) the locally Euclidean continuum
$X$ from an inverse system of locally finite open covers and the
finitary posets associated with them. In the end, the pointed $X$
was recovered from the said inverse system as a dense subset of
closed points of the system's projective limit space. Physically,
the inverse limit procedure was interpreted as the act of
topological refinement, as follows: as one employs finer and finer
(:`smaller' and `smaller') open sets to cover $X$ (:fincover
refinement), at the limit of infinite topological refinement, one
effectively ({\it ie}, modulo Hausdorff reflection
\cite{kopperman}) recovers the `classical' pointed topological
continuum $X$.

Back to our EG fintopos. In $\fintriad$ we represent the
aforementioned acts of topological refinement (:`topological
coarse graining') of the covering finsieves and their associated
finsheaves involved by `{\em differential geometric morphisms}'.
This is a new, finitary ADG-theoretic analogue of the fundamental
notion of {\em geometric morphism} in topos theory \cite{macmo}.
This definition of differential geometric morphism essentially
rests on a main result of Papatriantafillou
\cite{pap1,pap2,pap3,pap4} that a continuous map $f$ between
topological spaces (in our case, finitary poset substitutes of the
topological continuum) gives rise to a pair of maps (or,
categorically speaking, {\em adjoint functors}) $(f_{*},f^{*})$
that transfer backwards and forward (between the base finitary
posets) the differential structure encoded in the fintriads that
the finsheaves (of incidence algebras on Sorkin's finitary posets)
define. This is just one mathematical aspect of the {\em
functoriality} of our ADG-based constructions, but physically it
also supports our ADG-theoretic generalization of the Principle of
General Covariance (PGC) of the manifold based GR expressed in our
scheme via {\em natural transformations} between the relevant
functor (:structure sheaf) categories within $\ctriad$
\cite{malrap3,malrap4}. {\it In summa}, we will observe that
general covariance, as defined abstractly in $fcqv$-ADG-gravity,
is `preserved' under the said differential geometric morphisms
associated with Sorkin's acts of topological refinement.

More on the physics side, but quite heuristically, having
established that $\fintriad$ is a topos---a mathematical universe
in which geometry and logic are closely entwined \cite{macmo}, we
are poised to explore in the future deep connections between the
(quantum) {\em logic} and the (differential) {\em geometry} of the
vector and algebra finsheaves involved in the $fcqv$
Einstein-Lorentzian ADG-gravity. To this end, we could invoke
finite dimensional, irreducible (Hilbert space) matrix
representations $\hil$ of the incidence algebras dwelling in the
stalks of the finsheaves defining the fintriads in $\fintriad$,
and group them into {\em associated Hilbert finsheaves} $\hilb$
\cite{vas1,vas2,vas3,vas4}. Accordingly, via the associated
(:representation) sheaf functor \cite{macmo,mall1,vas4}, we can
organize the latter into the `{\em associated Hilbert fintopos}'
$\hilbert$. The upshot of these investigations could be the
identification, by using the abstract sheaf cohomological
machinery of ADG and the semantics of geometric prequantization
formulated {\it \`a la} ADG \cite{mall2,mall5,mall6,malrap2}, of
what we coin a `{\em quantum logical curvature}' form-like object
$\corv$ in $\fintriad$ and its representation Hilbert fintopos
$\hilbert$. $\corv$ has dual action and interpretation in
$(\fintriad ,\hilbert)$. From a differential geometric
(gravitational) standpoint (in $\fintriad$), $\corv$ marks the
well known obstruction to defining global (inertial) frames
(observers) in GR. This manifests itself in the fact that the
`curved' finsheaves of $\fintriad$ do not admit {\em global
elements}---{\it ie}, global sections. From a quantum-theoretic
(logical) one (in $\hilbert$), $\corv$ represents the equally well
known blockage to assigning values `globally' to (incompatible)
physical quantities in QT---the key feature of the `warped',
`twisted', contextual (:Boolean subalgebras' localized),
neorealist logic of quantum mechanics
\cite{buttish1,buttish2,buttish3,rawling}.

Accordingly, we envisage abstract `{\em sheaf cohomological
quantum commutation relations}' between certain characteristic
forms classifying the vector and algebra (fin)sheaves involved as
the {\it raison d'\^{e}tre} of the noted obstruction(s), similarly
to how in standard quantum mechanics the said inability to assign
global values to physical quantities is due to the Heisenberg
relations between incompatible observables such as position and
momentum. In fact, as we shall see in the sequel, the `forms'
defining the characteristic classes (of vector sheaves) in ADG,
and engaging into the abstract algebraic commutation relations to
be proposed, have analogous (albeit, abstract) interpretation as
`position' and `momentum' maps in the physical semantics of
geometrically prequantized ADG-field theory---in particular, as
the latter is applied to gravity (classical and/or quantum
ADG-gravity).

The paper is organized as follows: in the next section we recall
some basic properties of the abstract category $\ctriad$ of
differential triads as investigated recently by Papatriantafillou
\cite{pap1,pap2,pap3,pap4,pap5} and its application so far to
vacuum Einstein gravity \cite{mall3,malrap3,mall4}. With these in
hand, in the following section we present the category $\fintriad$
of fintriads involved in our $fcqv$-perspective on Lorentzian QG
\cite{malrap1,malrap2,malrap3,rap5,malrap4}, which is a full
subcategory of $\ctriad$, as an ET in the original sense due to
Lawvere and Tierney \cite{macmo}. Then, in section 4 we present
the same (fin)sheaf category as a GT by assuming a
Grothendieck-type of stance against Sorkin's locally finite poset
substitutes of continuous ({\it ie}, $\cont$-manifold) topologies.
This generalization rests essentially on identifying certain
covering coarse graining finsieves associated with Sorkin's
locally finite open covers of the original (spacetime) continuum
$X$ and on observing that they define a Grothendieck topology on
the poset category of open subsets of $X$. Under the categorical
prism of ADG as developed by Papatriantafillou, an offshoot of the
Grothendieck perspective on Sorkin is the categorical recasting of
topological refinement in Sorkin's inverse systems of finitary
poset substitutes of $X$ in terms of differential geometric
morphisms. This gives a differential geometric flavor to Sorkin's
originally purely topological acts of refinement, while the
finite, but more importantly the {\em infinite}, bicompleteness of
the fintopos $\fintriad$ secures the existence of a `classical'
continuum limit \cite{rapzap1,rapzap2,malrap2,malrap3} (triad) of
the coarse graining inverse system of fintriads in $\fintriad$. In
this respect, we observe that the abstract expression of the
Principle of General Covariance (PGC) of GR as the functoriality
of the ADG-vacuum Einstein gravitational dynamics with respect to
the structure sheaf $\struc$ of generalized coordinates is
preserved under differential geometric refinement. The paper
concludes with a fairly detailed, but largely heuristic and
tentative, discussion of four possible paths we could take along
current trends in `categorical quantum gravity' in order to
further develop our topos-theoretic scheme on $fcqv$-ADG-gravity.
More notably in this epilogue, we anticipate the aforesaid sheaf
cohomological quantum commutation relations, which may be regarded
as being responsible for the geometrico-logical obstructions
observed in $\fintriad$ and its associated (:representation)
Hilbert fintopos $\hilbert$. For the reader's convenience and
expository completeness, we have relegated the formal definitions
of an abstract elementary and an abstract Grothendieck topos to
two appendices at the end.

\section{Mathematical Formalities: The Category of Differential Triads and its Properties}

\subsection{ADG preliminaries: the physico-mathematical versatility and import of differential triads}

The principal notion in ADG is that of a {\em differential triad}
$\triad$. Let us briefly recall it, leaving more details to the
original sources \cite{mall1,mall2,mall4}.

We thus assume an in principle {\em arbitrary} topological space
$X$, which serves as the base localization space for the sheaves
to be involved in $\triad$. A differential triad then is thought
of as consisting of the following three ingredients:

\begin{enumerate}

\item A sheaf $\struc$ of unital, commutative and associative
$\K$-algebras ($\K=\R ,\com$) on $X$ called the {\em structure
sheaf} of generalized arithmetics in the
theory.\footnote{`Coordinates' or `coefficient functions' are
synonyms to `arithmetics'.}

\item A sheaf $\Omg$ of $\K$-vector spaces over $X$, which is an
$\struc(U)$-module ($\forall$ open $U$ in $X$).

\item A $\K$-linear and Leibnizian relative to $\struc$ map (:sheaf morphism) $\partial$ between $\struc$ and $\Omg$,

\begin{equation}\label{eq0}
\partial :~\struc\mapto\omg
\end{equation}

\noindent which is the archetypical paradigm of a (flat)
$\struc${\em -connection} in ADG \cite{mall-3,mall-2}.

\end{enumerate}

\noindent {\it In toto}, a differential triad is represented by
the triplet:

\begin{equation}\label{eq1}
\triad :=(\struc_{X} ,\partial ,\Omg_{X})
\end{equation}

\noindent Or, omitting the base topological space $X$ (as we shall
often do in the sequel), $\triad=(\struc ,\partial ,\Omg)$.

A couple of additional technical remarks on differential triads
are due here for expository completeness:

\begin{itemize}

\item The constant sheaf $\mathbf{K}$ of scalars $\K$ is naturally injected into $\struc$:
$\mathbf{K}\stackrel{\subset}{\hookrightarrow}{\struc}$.

\item In general, a {\em vector sheaf} $\modl$ in ADG is defined as a locally free $\struc$-module of finite rank $n$,
by which it is meant that, locally in $X$ (:$\forall$ open
$U\subseteq X$), $\modl$ is expressible as a finite power (or
equivalently, a finite Whitney sum) of $\struc$:
$\modl(U)\simeq(\struc(U))^{n}\equiv\struc^{n}(U)$, with $n$ a
positive integer called the {\em rank} of the sheaf, and
$\modl(U)\equiv\Gamma(U,\modl)$ the space of local sections of
$\modl$ over $U$. It is also assumed that such a vector sheaf
$\modl$ is the dual of the $\struc$-module sheaf $\Omg$ appearing
in the triad in (\ref{eq1}), {\it ie},
$\modl^{*}=\Omg(\equiv\Omg^{1})={\Hom}_{\struc}(\modl ,\struc)$.

\item As it has been repeatedly highlighted in thorough investigations on various properties
and in numerous (physical) applications of differential triads
\cite{pap1,pap2,pap3,pap4,pap5,malrap2,malrap3,malrap4,rap5}, the
latter generalize differential (:$\smooth$-smooth) manifolds, and,
{\it in extenso}, ADG abstracts from and generalizes the usual
differential geometry of smooth manifolds---{\it ie}, the standard
Differential Calculus on manifolds, which we have hitherto coined
{\em Classical Differential Geometry} (CDG)
\cite{malrap1,malrap2,malrap3,malrap4,rap5}. Indeed, CDG may be
thought of as a `reduction' ({\it ie}, a particular instance) of
ADG, when one assumes $\smooth_{X}$---the usual sheaf of germs of
smooth ($\K$-valued) functions on $X$---as structure sheaf in the
theory. In this particular case, $X$ is a smooth manifold $M$,
while the $\Omg$ involved in the corresponding `classical'
differential triad is the usual sheaf of germs of local
differential $1$-forms on (:cotangent to) $M$.\footnote{{\it In
summa}, when $\struc\equiv\smooth_{X}$, $X$ is a differential
manifold $M$, $\modl$ is the tangent bundle $TM$ of smooth vector
fields on $M$, while $\Omg$ the cotangent bundle $T^{*}M$ of
smooth $1$-forms on $M$, which is the dual to $TM$. Note here that
in the purely algebraic (:sheaf-theoretic) ADG, there are {\it a
priori} no such central CDG-notions as {\em base manifold}, {\em
(co)tangent space} (to it), {\em (co)tangent bundle} {\it etc.}
ADG deals directly with the algebraic structure of the sheaves
involved (:the algebraic relations between their sections),
without recourse to (or dependence on) a  background geometrical
`continuum space' (:manifold) for its differential geometric
support. In this sense ADG is completely Calculus-free
\cite{mall1,mall2}.} However, and this is the versatility of ADG,
one need {\em not} restrict oneself to $\struc\equiv\smooth_{X}$
hence also to the usual theory (CDG on manifolds). Instead, one
can assume `non-classical' structure sheaves that may appear to be
`exotic' ({\it eg}, non-functional) or very `pathological' ({\it
eg}, singular) from the `classical' vantage of the featureless
smooth continuum and the CDG it supports, provided of course that
these algebra sheaves of generalized arithmetics furnish one with
a differential operator $\partial$ with which one can set up a
triad in the first place. Parenthetically, an example of the said
`exotic', non-functional structure sheaves that have been used in
numerous applications of ADG to gravity are sheaves of {\em
differential incidence algebras of finitary posets}
\cite{malrap1,malrap2,malrap3,malrap4,rap5}.\footnote{They are
also due to appear in the sequel.} At the same time, as very
`pathological', `ultra-singular' structure sheaves, one may regard
sheaves of Rosinger's {\em differential algebras of non-linear
generalized functions} (:distributions), hosting singularities of
all kinds densely in the underlying $X$. These too have so far
been successfully applied to GR
\cite{malros1,malros2,mall3,malros3,mall7,mall11,malrap4,rap5,mall4}.

\end{itemize}

\paragraph{Connection, curvature, field and (vacuum) Einstein ADG-gravity.}
Differential triads are versatile enough to support such key
differential geometric concepts as {\em connection} and {\em
curvature}. They can also accommodate central GR notions such as
the (vacuum) {\em gravitational field} and the (vacuum) Einstein
differential equations that it obeys. For expository completeness,
but {\it en passant}, let us recall these notions from
\cite{mall1,mall2,mall3,malrap1,malrap2,malrap3,rap5,malrap4}:

\medskip

\noindent {\bf $\struc$-connections:} An $\struc$-connection
$\conn$ is a (`curved') generalization of the (flat) $\partial$ in
(\ref{eq0}) and its corresponding differential triad (\ref{eq1}).
It too is defined as a $\mathbf{K}$-linear and Leibnizian sheaf
morphism, as follows

\begin{equation}\label{eqx1}
\conn :~\modl\mapto\Omg(\modl)\equiv\modl\otimes_{\struc}\Omg\cong
\Omg\otimes_{\struc}\modl
\end{equation}

\medskip

\noindent {\bf Curvature of an $\struc$-connection:} With $\conn$
in hand, we can define its curvature $\curv(\conn)$
diagrammatically as follows

\begin{equation}\label{eqx7}
\Vtriangle[\modl`\Omg^{1}(\modl)\equiv\modl\otimes_{\struc}\Omg^{1}`\Omg^{2}(\modl)\equiv\modl\otimes_{\struc}
\Omg^{2};\conn`\curv\equiv\conn^{1}\circ\conn`\conn^{1}]
\end{equation}

\noindent for a higher-order prolongation $\conn^{2}$ of
$\conn(\equiv\conn^{1})$. One can then define the Ricci curvature
$\ric$, as well as its trace---the Ricci scalar $\ricci$.
$\curv(\conn)$, unlike $\conn$ which is only a constant sheaf
$\mathbf{K}$-morphism, is an $\struc$-morphism, {\it alias}, an
$\otimes_{\struc}$-tensor (with $\otimes_{\struc}$ the usual
homological tensor product functor).

\medskip

\noindent {\bf ADG-field:} In ADG, the pair

\begin{equation}\label{eqx3}
(\modl ,\conn)
\end{equation}

\noindent namely, a connection $\conn$ on a vector sheaf $\modl$,
is generically called a {\em field}. $\modl$ is thought of as the
{\em carrier space} of the connection, and $\conn$ acts on its
(local) sections. Note that there is no base (spacetime) manifold
whatsoever supporting the ADG-field, so that the latter is a
manifestly ({\it ie}, by definition/construction) background
manifold independent entity.

\medskip

\noindent {\bf Vacuum Einstein equations:} The {\em vacuum
ADG-gravitational field} is defined to be the field $(\modl
,\conn)$ whose connection part has a Ricci scalar curvature
$\ricci(\conn)$ satisfying the vacuum Einstein equations

\begin{equation}\label{eqx4}
\ricci(\modl)=0
\end{equation}

\noindent on the carrier sheaf $\modl$. (\ref{eqx4}) can be
derived from the variation of an Einstein-Hilbert action
functional $\eh$ on the affine space $\sconn_{\struc}(\modl)$ of
$\struc$-connections $\conn$ on $\modl$. In (vacuum) ADG-gravity,
the sole dynamical variable is the gravitational connection
$\conn$, thus the theory has been coined `{\em pure gauge theory}'
and the formalism supporting it `{\em half-order formalism}'
\cite{malrap3,rap5,malrap4}.

\medskip

\noindent Overall, applications to gravity (classical or quantum)
aside for the moment, and in view of the categorical perspective
that we wish to adopt in the present paper, perhaps the most
important remark that can be made about differential triads is
that {\em they form a category} $\ctriad$, in which, as befits the
aforementioned generalization of CDG by ADG, the category $\man$
of differential manifolds is embedded \cite{pap1}. Thus, in the
next subsection we recall certain basic categorical features of
$\ctriad$ from \cite{pap1,pap2,pap3,pap4,pap5}, which will prove
to be very useful in our topos-theoretic musings subsequently.

\subsection{The categorical perspective on differential triads}

As noted above, in the present subsection we draw material and
results from Papatriantafillou's inspired work on the properties
of $\ctriad$, ultimately with an eye towards revealing its true
topos-theoretic colors. Thus, below we itemize certain basic
features of $\ctriad$, with potential topos-theoretic significance
to us as we shall see in the next section, as were originally
exposed in \cite{pap1,pap2,pap3,pap4,pap5}. For more technical
details, such as formal definitions, relevant proofs, {\it etc.},
the reader can refer to those original papers.

However, before we discuss the properties of $\ctriad$, we must
first emphasize that it is indeed a category proper. {\em Objects}
in $\ctriad$ are differential triads, while {\em arrows} between
them are {\em differential triad morphisms}. Let us recall briefly
from \cite{pap1,pap4,pap5} what the latter stand for.

\paragraph{Enter geometric morphisms.} To discuss morphisms of differential triads, we first bring forth from \cite{macmo}
a pair of (covariant) adjoint functors between sheaf categories
that are going to be of great import in the sequel.

Let $X$, $Y$ be topological spaces, and $\sh_{X}$, $\sh_{Y}$ sheaf
categories over them. Then, a continuous map $f:~X\mapto Y$
induces a pair $\geomorph_{f}=(f_{*},f^{*})$ of (covariant)
adjoint functors between $\sh_{X}$ and $\sh_{Y}$
($f_{*}:~\sh_{X}\mapto\sh_{Y}$, $f^{*}:~\sh_{Y}\mapto\sh_{X}$)
called {\em push-out} (:direct image) and {\em pull-back}
(:inverse image), respectively. In topos-theoretic parlance, such
a pair of adjoint functors is known as a {\em geometric morphism}
\cite{macmo}.

With $\geomorph$ in hand, we are in a position to define
differential triad morphisms. Let
$\triad_{X}=(\struc_{X},\partial_{X},\Omg_{X})$ and
$\triad_{Y}=(\struc_{Y},\partial_{Y},\Omg_{Y})$ be differential
triads over the aforesaid topological spaces. Then, like the
triads themselves, a {\em morphism} $\morph$ between them is a
triplet of maps $\morph=(f,f_{A},f_{\Omg})$ having the following
four properties relative to $\geomorph_{f}$:

\begin{enumerate}

\item the map $f:~X\mapto Y$ is continuous, as set by $\geomorph_{f}$;

\item the map $f_{\struc}:~\struc_{Y}\mapto f_{*}(\struc_{X})$ is
a morphism of sheaves of $\mathbb{K}$-algebras over $Y$, which
preserves the respective algebras' unit elements ({\it ie},
$f_{\struc}(1)=1$);

\item the map $f_{\Omg}:~\Omg_{Y}\mapto f_{*}(\Omg_{X})$ is a morphism of sheaves of $\mathbb{K}$-vector
spaces over $Y$, with
$f_{\Omg}(\alpha\omega)=f_{\struc}(\alpha)f_{\Omg}(\omega),~\forall
(\alpha ,\omega)\in\struc_{Y}\times_{Y}\Omg_{Y}$; and finally,

\item with respect to the $\mathbf{K}$-linear, Leibnizian sheaf morphism $\partial$ in the respective triads,
the following diagram is commutative:

\[
\bfig
\putsquare<1`1`1`1;600`600>(600,600)[\struc_{Y}`\Omg_{Y}`f_{*}(\struc_{X})`f_{*}(\Omg_{X});
\partial_{Y}`f_{\struc}`f_{\Omg}`f_{*}(\partial_{X})]
\efig
\]

\noindent reading: $f_{\Omg}\circ
\partial_{Y}=f_{*}(\partial_{X})\circ f_{\struc}$.

\end{enumerate}

\noindent To complete the argument that $\ctriad$ is a true
category, we note that for each triad $\triad_{X}$ there is an
{\em identity morphism}
$id_{\triad_{X}}:=(id_{x},id_{\struc},id_{\Omg})$ defined by the
corresponding identity maps of the spaces involved in the triad.
There is also an {\em associative composition law} (:product)
between triad morphisms \cite{pap1,pap4,pap5}, making thus
$\ctriad$ an arrow (:triad morphism) semigroup, complete with
identities (:units)---one for every triad object in it.

Here, we would like to make some auxiliary and clarifying remarks
about $\ctriad$ that will prove to be helpful subsequently:

\begin{itemize}

\item For a given base space $X$, the collection
$\{(\triad_{X},\morph_{X}:=(id_{X},f_{\struc},f_{\Omg}))\}$
constitutes a subcategory of $\ctriad$, symbolized as
$\ctriad_{X}$.

\item In general, differential triad morphisms are thought of as {\em maps
that preserve the purely algebraic (:sheaf-theoretic) differential
(geometric) structure or `mechanism' encoded in every triad}. They
are {\em abstract differentiable maps}, generalizing in many ways
the usual smooth ones between differential manifolds in $\man$.

\item Indeed, following the classical (CDG) jargon, in ADG we say that a
continuous map $f:~X\mapto Y$ is {\em differentiable}, if it can
be completed to a differential triad morphism. Such
continuity-to-differentiability completions of maps, in striking
contradistinction to the case of $\man$ and CDG, are always
feasible in $\ctriad$ and ADG, as the following two results show:

\item If $X$ and $Y$ are topological spaces as before, $f:~X\mapto
Y$ continuous, and $X$ carries a differential triad $\triad_{X}$,
the push-out $f_{*}$ induces a differential triad on $Y$. Vice
versa, if $Y$ carries a differential triad, the pull-back $f^{*}$
endows $X$ with a differential triad. Furthermore, these so-called
`{\em final and initial differential structures}' respectively,
satisfy certain {\em universal mapping} relations that promote $f$
to a {\em differentiable} map in the sense above---{\it ie}, they
complete it to a triad morphism
\cite{pap3,pap4,pap5}.\footnote{Furthermore, as shown in
\cite{pap3,pap4,pap5}, the composition of a differentiable map
with a continuous one also becomes differentiable in the sense
above.} This is in glaring contrast with the usual situation in
$\man$, whereby if $X$ is a smooth manifold equipped with an atlas
$\mathcal{A}$, while $Y$ is just a topological space, one cannot
push-forward $\mathcal{A}$ by $f_{*}$ in order to make $Y$ a
differential manifold and in the process turn $f$ into a smooth
map. Similarly for the reverse scenario in which $Y$ is a
differential manifold charted by a smooth atlas $\mathcal{B}$, and
$X$ simply a topological space: $f^{*}$ cannot `pull back'
$\mathcal{B}$ on $X$ thus promote the latter into a smooth
space.\footnote{Results like this have prompted workers in ADG to
develop, as an extension of the usual `{\em differential geometry
of smooth manifolds}' (:CDG in $\man$), what one one could coin
`{\em the differential geometry of topological spaces}' (ADG in
$\ctriad$). This possibility hinges on the following
`Calculus-reversal' observed in ADG and noted above: as in the
usual CDG on manifolds `{\em differentiability implies
continuity}' (:a smooth map is automatically continuous), in ADG
the converse is also possible; namely, that `{\em continuity
implies differentiability!}' (:a continuous map can become
differentiable).}

\item Much in the same fashion, if $X$ is a differential manifold and
$\sim$ an (arbitrary) equivalence relation on it, the moduli space
$X/\!\!\sim$ does not inherit, via the (possibly continuous)
canonical projection $f_{\sim}:~X\mapto X/\!\!\sim$, the usual
differential structure of $X$ (and, accordingly, the possibly
continuous surjection $f$ does {\em not} become differentiable in
the process). This is not the case in $\ctriad$; whereby, when the
base space of a differential triad is modded-out by an equivalence
relation, the resulting quotient space inherits the original
triad's structure ({\it ie}, it becomes itself a differential
triad), and in the process $f_{\sim}$ becomes a triad morphism
\cite{pap3,pap4,pap5}. This particular example of the versatility
of $\ctriad$ (and ADG!), as contrasted against the `rigidity' of
$\man$ (and CDG!), has been exploited numerous times in the past,
especially in the finitary case of Sorkin
\cite{malrap1,malrap2,malrap3,rap5,malrap4}. We shall exploit it
again later in this paper.

\end{itemize}

\noindent After these telling preliminaries, we return to discuss
the categorical properties of $\ctriad$ that will be of potential
topos-theoretic significance in the sequel. Once again, we itemize
them, commenting briefly on every item:

\begin{itemize}

\item $\ctriad$ {\em is bicomplete}. This means that $\ctriad$ is
{\em closed} under both {\em inverse} ({\it alias}, projective)
and {\em direct} ({\it alias}, inductive) limits of differential
triads \cite{pap2,pap5}. In particular, it is closed under {\em
finite} limits (projective) and colimits (inductive)---{\it ie},
it is {\em finitely bicomplete}.\footnote{The synonyms
`co-complete' or `co-closed' are often used instead of
`bicomplete'.}

\item $\ctriad$ {\em has canonical subobjects}. As it has been
shown in detail in \cite{pap4,pap1,pap5}, ``{\em every subset of
the base space of a differential triad defines a differential
triad, which is a subobject of the former}''.\footnote{Excerpt
from the abstract of \cite{pap4}.} On the other hand, $\man$
manifestly lacks this property, since it is plain that an
arbitrary subset of a differential manifold is not itself a
manifold.

\item $\ctriad$ {\em has finite products}. In
\cite{pap4,pap1,pap5} it is also shown that there are {\em finite
cartesian products} of differential triads in $\ctriad$.

\item $\ctriad$ {\em has an exponential structure}. This means
that given any two differential triads $\triad
,\triad^{'}\in\ctriad$, one can form the collection
$\triad^{'\triad}$ of all triad morphisms in $\ctriad$ from
$\triad$ to $\triad^{'}$. Common in categorical notation is the
alternative designation of $\triad^{'\triad}$ by $\Hom(\triad
,\triad^{'})$ (:`hom-sets of triad morphisms'). In addition, the
exponential is supposed to effectuate canonical isomorphisms
relative to the aforementioned (cartesian) product in $\ctriad$.
Thus, for $\triad$, $\triad^{'}$ as above and $\triad^{''}$ any
other triad in $\cat$: $\Hom(\triad^{''}\times \triad,
\triad^{'})\simeq\Hom(\triad^{''},\triad^{'\triad})$ (or
equivalently: $\triad^{'\triad^{''}\!\!\times\triad}\simeq
(\triad^{'\triad})^{\triad^{''}}$).

\end{itemize}

\section{The Category of Fintriads as an ET}

The high-point of the present section is that we show that the
category $\fintriad$ of fintriads, which is a subcategory of
$\ctriad$, is an ET \cite{macmo}. Before we do this however, let
us first recall {\it en passant} the finitary perspective on
continuous (spacetime) topology as originally championed by Sorkin
\cite{sork0} and then extended by the sheaf-theoretic ADG-means to
the differential geometric realm, with numerous physical
applications to discrete and quantum spacetime structure (:causets
and qausets) \cite{rapzap1,rapzap2,rap1,rap2}, vacuum
Einstein-Lorentzian gravity (classical and quantum)
\cite{malrap1,malrap2,malrap3}, and gravitational singularities
\cite{rap5,malrap4}.

\subsection{`Finatarities' revisited}

Below, we give a short, step-by-step historical anadromy to the
development of the finitary spacetime and gravity program by
ADG-theoretic means, isolating and highlighting the points that
are going to be relevant to our ADG {\it cum} topos-theoretic
efforts subsequently.\footnote{For more details on what follows,
the reader is referred to the aforementioned papers on the
finitary ADG-based approach to spacetime and gravity.} The account
concludes with the arrival at the category $\fintriad$ of
fintriads, which we present as an ET proper in the following two
subsections:

\begin{itemize}

\item {\bf Finitary poset substitutes of topological manifolds.}
In \cite{sork0}, Sorkin commenced the finitary spacetime program
solely with topology in mind. Namely, he substituted the usual
continuous ($\cont$) topology of an open and bounded region $X$ of
the spacetime manifold $M$ by a poset $P_{i}$ relative to a
locally finite open covering $\gauge_{i}$\footnote{The index `$i$'
will be explained shortly.} of $X$. He arrived at $P_{i}$, which
is a $T_{0}$-topological space in its own right, by factoring out
$X$ by the following {\em equivalence relation} between $X$'s
points:

\begin{equation}\label{eq2}
\begin{array}{c}
x\stackrel{\gauge_{i}}{\sim}y\Leftrightarrow
\Lambda(x)|_{\gauge_{i}}=\Lambda(y)|_{\gauge_{i}},~\forall x,y\in
X;\cr P_{i}:=X/\stackrel{\gauge_{i}}{\sim}
\end{array}
\end{equation}

\noindent with $\Lambda(x)|_{\gauge_{i}}:=\bigcap\{
U\in\gauge_{i}:~ x\in U\}$ the smallest open set in $\gauge_{i}$
(or equivalently, in the subtopology $\tau_{i}$ of $X$ generated
by taking arbitrary unions of finite intersections of the covering
open sets in $\gauge_{i}$) containing $x$.
$\Lambda(x)|_{\gauge_{i}}$ is otherwise known as the
Alexandrov-\v{C}ech {\em nerve} of $x$ relative to the open
covering $\gauge_{i}$ \cite{rapzap1}. The `points' of $P_{i}$ are
equivalence classes (:nerves) of $X$'s points, partially ordered
by set-theoretic inclusion `$\subseteq$', with the said
equivalence relation being interpreted as `indistinguishability'
of points relative to our `coarse measurements' in $\gauge_{i}$.
That is to say, two points (:`events') of $X$ in the same class
cannot be distinguished (or `separated', topologically speaking)
by the covering open sets (:our `coarse measurements') in
$\gauge_{i}$.

Sorkin's scenario for approximating (:`substituting') the locally
Euclidean (:continuum) topology of $X$ by the finitary topological
posets (:fintoposets) $P_{i}$ hinges on the fact that the latter
constitute an {\em inverse} ({\it alias}, {\em projective}) system
$\inv$ relative to a {\em topological refinement net}
$I\equiv\invf :=\{(\gauge_{i},\preceq\}_{i\in I}$ of the fincovers
of $X$. Here, the partial order $\gauge_{i}\preceq\gauge_{j}$ is
interpreted as follows: `{\em the covering $\gauge_{j}$ (resp.
$\gauge_{i}$) is finer (resp. coarser) than $\gauge_{i}$ (resp.
$\gauge_{j}$)}'. Equivalently, $\gauge_{j}$ is a refinement of
$\gauge_{i}$, and thus the latter is a subcover of the former.
Correspondingly, $\tau_{i}$ is a subtopology of $\tau_{j}$.
Henceforth, the net $I$ will be treated as an index-set labelling
the open coverings of $X$ and the corresponding fintoposets.
However, we may use the symbols $\invf$ and $I$ interchangeably,
hopefully without causing any confusion. Parenthetically, and with
an eye towards our subsequent topos-theoretic labors in the light
of $\ctriad$, it is worth pointing out here that $\inv$ can be
described as an $I$-indexed family of pentads:

\begin{equation}\label{eq3}
\inv:=\{(X,f_{i},P_{i},f_{j},P_{j},f_{ji})\} ,~(i\preceq j\in I)
\end{equation}

\noindent whereby, $F_{i}$ (resp. $F_{j}$) is a continuous
surjection (:projection map) from $X$ to $P_{i}$ (resp. $P_{j}$),
while $f_{ji}$ a continuous fintoposet morphism from $P_{j}$ to
$P_{i}$ corresponding to the act of topological refinement when
one refines the open cover $\gauge_{i}$ to $\gauge_{j}$. {\it En
passant}, we note that the epithet `continuous' for $f_{ji}$ above
pertains to the fact that one can assign a `natural'
topology---the so-called Sorkin {\em lower-set} or {\em
sieve-topology}---to the $P_{i}$s, whereby an open set is of the
form $\mathcal{O}(x):=\{ y\in P_{i}:~y\mapto x\}$, with `$\mapto$'
the partial order relation in $P_{i}$. Basic open sets for the
Sorkin topology are defined via the {\em links} or {\em covering}
(`immediate arrow') relations in (the Hasse diagram of) $P_{i}$:
$\mathcal{O}_{B}(x):=\{ y\in P_{i}:~ (y\mapto x)\wedge(\nexists
z\in P_{i}:~y\mapto z\mapto x)\}$. Then, $f_{ji}$ is a monotone
(:partial order preserving) surjection from $P_{j}$ to $P_{i}$,
hence it is continuous with respect to the said Sorkin
sieve-topology. Accordingly, the arrow $x\mapto y$ can be
literally interpreted as the convergence of the constant sequence
$(x)$ to $y$ in the Sorkin topology \cite{sork0}. This
sieve-topology of Sorkin will prove to be very important for our
topos-theoretic (and especially the GT) musings in the
sequel.\footnote{See next section and appendix \ref{app2}.}

From \cite{sork0} we also recall that there is a {\em universal
mapping condition} obeyed by the triplets $(F_{i},F_{j},f_{ji})$
of continuous surjective maps in $\inv$, which looks
diagrammatically as follows

\begin{equation}\label{eq4}
\qtriangle[X`P_{j}`P_{i};F_{j}`F_{i}`f_{ji}]
\end{equation}

\noindent and reads: $F_{i}=f_{ji}\circ F_{j}$. That is, the
system $(F_{i})_{i\in I}$ of canonical projections of $X$ onto the
fintoposets is `{\em universal}' as far as maps between
$T_{0}$-spaces are concerned, with $f_{ji}$ the {\em unique}
map---itself a partial order preserving (monotone) surjection of
$P_{j}$ onto $P_{i}$---mediating between the continuous
projections $F_{i}$ and $F_{j}$ of $X$ onto the
$T_{0}$-fintoposets $P_{i}$ and $P_{j}$, respectively.

Then, the central result in \cite{sork0}---the one that qualifies
the fintoposets as genuine finitary approximations of the
continuous topology of the $\cont$-manifold $X$---is that, thanks
to the universal mapping property that the $P_{i}$s enjoy, at the
projective limit of infinite topological refinement (:coarse
graining) of the underlying coverings in $I$, $\inv$
effectively\footnote{That is, modulo Hausdorff reflection
\cite{kopperman}.} yields back the original topological continuum
$X$ (up to homeomorphism). Formally, one writes:

\begin{equation}\label{eq5}
\underset{\infty\leftarrow
j}{\lim}f_{ji}(P_{i})=:P_{\infty}\underset{\mathrm{homeo.}}{\overset{F_{\infty}}{\leftrightarrows}}
X~\mathrm{(modulo~Hausdorff~reflection)}
\end{equation}

\noindent Let it be noted here that this universal mapping
property of the maps between the $T_{0}$-fintoposets above is
completely analogous to the one possessed by the differential
triad morphisms ({\it eg}, the push-outs and pull-backs along
continuous maps between the triads' base topological spaces)
\cite{pap3,pap4,pap5} mentioned earlier in 2.2. In fact, shortly,
when we discuss fintriads and their inverse limits, the ideas of
Sorkin and Papatriantafillou will appear to be tailor-cut for each
other; albeit, with the ADG-based work of Papatriantafillou adding
an important {\em differential} geometric slant to Sorkin's
originally purely topological considerations.

\item {\bf Incidence algebras of $T_{0}$-posets.} One can use a
discrete version of Gel'fand duality to represent the fintoposets
$P_{i}$ above algebraically, as so-called {\em incidence algebras}
(write $\omg_{i}(P_{i})$, and read `{\em the incidence algebra
$\omg_{i}$ of the fintoposet} $P_{i}$') \cite{rapzap1,rapzap2}.
The correspondence $P_{i}\mapto\omg_{i}$ is manifestly {\em
functorial},\footnote{The reverse correspondence $\omg_{i}\mapto
P_{i}$ having been coined {\em Gel'fand spatialization}
\cite{zap0,rapzap1,rapzap2}.} especially when one regards the
$P_{i}$s as graded simplicial complexes having for simplices the
aforementioned \v{C}ech-Alexandrov nerves
\cite{rapzap1,rapzap2,zap1}.

The $\omg_{i}$s, being (categorically) dual objects to the
`discrete' homological (:simplicial) $P_{i}$s, may be regarded as
`{\em $\Z_{+}$-graded discrete differential
$\mathbb{K}$-algebras}'---reticular cohomological analogues of the
usual spaces (:modules) of (smooth) differential forms on the
manifold $X$ in focus \cite{rapzap1,rapzap2}. Indeed, they were
seen to be `{\em discrete differential manifolds}' (in the sense
of \cite{dimu1,dimu2,dimu4}), as follows

\begin{equation}\label{eq6}
\omg_{i}=\bigoplus_{p\in\Z_{+}}\omg^{p}_{i}=
\stackrel{{\mathcal{A}}_{i}}{\overbrace{\omg^{0}_{i}}}\oplus
\stackrel{{\mathcal{R}}_{i}}{\overbrace{\omg^{1}_{i}\oplus
\omg^{2}_{i}\oplus\ldots}}\equiv{\mathcal{A}}_{i}\oplus{\mathcal{R}}_{i}
\end{equation}

\noindent where ${\mathcal{R}}_{i}$ is a $\Z_{+}$-graded
${\mathcal{A}}_{i}$-bimodule of (exterior, real or complex)
differential form-like entities
$\omg^{p}_{i}~(p\geq1)$,\footnote{In (\ref{eq6}) above,
${\mathcal{A}}_{i}\equiv\omg^{0}_{i}$ is a commutative subalgebra
of $\omg_{i}$ called {\em the algebra of coordinate functions in}
$\omg_{i}$, while
${\mathcal{R}}_{i}\equiv\bigoplus_{i}^{p\geq1}\omg^{p}_{i}$ a
linear subspace of $\omg_{i}$ called {\em the module of
differentials over} ${\mathcal{A}}_{i}$. The elements of each
linear subspace $\omg^{p}_{i}$ of $\omg_{i}$ in
${\mathcal{R}}_{i}$ have been regarded as `discrete' analogues of
the usual smooth differential $p$-forms teeming the usual
(cotangent bundle over the) smooth manifold
\cite{rapzap1,rapzap2}.} related within each $\omg_{i}$ by
nilpotent Cartan-K\"{a}hler-type of (exterior) differential
operators $d^{p}_{i}:~\omg^{p}_{i}\mapto\omg^{p+1}_{i}$, with
$d^{0}_{i}\equiv\partial_{i}:~\omg^{0}\mapto\omg^{1}$ the finitary
analogue of the standard derivation $\partial$ in (\ref{eq1}) and
$d^{p}_{i}:~\omg^{p}_{i}\mapto\omg^{p+1}_{i}~(p\geq1)$ its higher
order (grade or degree) prolongations.

Plainly, it is tacitly assumed here that the locally Euclidean
$X$, apart from the continuous ($\cont$) topology, also carries
the usual differential (:$\smooth$-smooth) structure, which in
turn the $\omg_{i}$s can be thought of as approximating
`discretely' (`finitarily'). Thus, the cohomological $\omg_{i}$s
too are seen to comprise an inverse system $\invomg$ relative to
the aforesaid topological refinement net of fincovers of $X$
\cite{malrap2,malrap3,rap5,malrap4}.

\item {\bf Finsheaves of incidence algebras: fintriads.} The key
observation in arriving at {\em finsheaves} \cite{rap2} of
incidence algebras is that by construction ({\it ie}, by the
aforesaid method of Gel'fand spatialization) the map
$\omg_{i}\mapto P_{i}$ is a {\em local homeomorphism}, {\it
alias}, a {\em sheaf} \cite{macmo,mall1,mall2}. Thus, finsheaves
$\Omg_{i}(P_{i})$ of incidence algebras over Sorkin's fintoposets
were born \cite{malrap1}. Moreover, since the $\omg_{i}$s carry
not only topological, but also {\em differential geometric}
structure as noted above, the $\Omg_{i}$s may be thought of as
finitary analogues (`approximations') of the `classical'
(:$\smooth$-smooth) differential triad $\triad_{\infty}$ supported
by the differential manifold $X$. They are coined `{\em
fintriads}' and they are fittingly symbolized by $\triad_{i}$
\cite{malrap2,malrap3,rap5,malrap4}.

There are actually two ways to arrive at $\triad_{i}$s---one
`indirect' and `constructive', the other `direct' and `inductive':

\begin{enumerate}

\item The `indirect-constructive' way is the one briefly described
above, namely, by first obtaining the $P_{i}$s from Sorkin's
factorization algorithm, then by defining the corresponding
$\omg_{i}$s and suitably topologizing them in a `discrete'
Gel'fand representation (:duality) fashion, and finally, by
defining finsheaves of the latter (regarded as discrete
differential algebras) over the former. This is the path we
followed originally in our work \cite{malrap1,malrap2,malrap3}.

\item The `direct-inductive' way goes as follows \cite{rap5,malrap4}: one simply
starts with the `classical' smooth differential triad
$\triad_{\infty}$ on the (differential) manifold $X$ and, by
calling forth Papatriantafillou's push-out/pull-back results in
\cite{pap3,pap5} that we mentioned back in 2.2, one induces the
usual differential structure, via the push-out $F_{i*}$ of the
continuous surjection $F_{i}:~X\mapto P_{i}$ in (\ref{eq4}) above,
from $X$ to the $\stackrel{\gauge_{i}}{\sim}$-moduli space
$P_{i}$. In the process, $F_{i}$ becomes differentiable---{\it
ie}, it lifts to a triad morphism
$\morph_{i}:~\triad_{\infty}\mapto\triad_{i}$. Incidentally, from
Papatriantafillou's results \cite{pap3,pap4,pap5} it follows that
the continuous surjection $f_{ji}$ in (\ref{eq4}) is also promoted
to a fintriad morphism $\morph_{ji}:~\triad_{j}\mapto\triad_{i}$.

\end{enumerate}

\noindent From \cite{malrap1,malrap2,malrap3,rap5,malrap4} we draw
that a fintriad can be symbolized as
$\triad_{i}=(\struc_{i},\partial_{i},\Omg_{i})$, where
$\struc_{i}$ is a unital, abelian, associative algebra (structure)
sheaf whose stalks are inhabited by elements (:coordinate
function-like entities) of ${\mathcal{A}}_{i}$ in (\ref{eq6}),
while $\Omg_{i}$ is an $\struc_{i}$-bimodule with elements
(:differential form-like entities) of ${\mathcal{R}}_{i}$ in
(\ref{eq6}) dwelling in its fibers.

\item {\bf The category $\fintriad$ and its completeness in $\ctriad$.} Finally,
we can organize the $\triad_{i}$s into the category $\fintriad$ of
fintriads. Objects in $\fintriad$ are the said fintriads, while
arrows between them fintriad morphisms. It is easy to see that
$\fintriad$ is a {\em full subcategory} of $\ctriad$ \cite{macmo}.

An important result about $\fintriad$ is that it is finitely
complete in itself, and infinitely complete in $\ctriad$---{\it
ie}, it is closed under finite projective limits, and closed in
$\ctriad$ under infinite inverse limits. This is a corollary
result which derives---also bearing in mind Sorkin's inverse limit
result about $\inv$ (\ref{eq5}), as well as the universal mapping
properties observed in both Sorkin's scheme \cite{sork0} and in
$\ctriad$ \cite{pap3,pap4,pap5}---from the following theorem
proved by Papatriantafillou in \cite{pap2,pap5}:\footnote{Quoting
the author almost verbatim from \cite{pap2} (theorem 4.4), with
slight changes in notation and language to suit our finitary
considerations here.}

\medskip

\noindent {\bf Theorem:} Let
$[\triad_{i}=(\struc_{i},\partial_{i},\Omg_{i});\morph_{ji}=(f_{ji},f_{ji\struc},f_{ji\Omg})]_{i\preceq
j\in I}$ be a projective system in $\fintriad\subset\ctriad$ and
let $P_{i}$ be the base space of each $\triad_{i}$, with
$\inv=(P_{i},f_{ji})$ their inverse system considered above. There
is a differential triad $\triad_{\infty}$ over the inverse limit
space $X\stackrel{\mathrm{homeo.}}{\simeq}P_{\infty}$ in
(\ref{eq5}), satisfying the universal property of the projective
limit in $\ctriad$.

\medskip

Dually, the same would hold for an {\em inductive system} of
differential triads over an inductive system of base spaces and
their direct limit space \cite{pap2,pap5}. Here, in connection
with Sorkin's `finitarities' \cite{sork0}, we happen to be
interested only in the projective (inverse) limit case, but
$\fintriad$ is also co-complete in $\ctriad$. In fact, it is
noteworthy here that {\em inductive} systems of fintoposets (as
base spaces) were originally employed in \cite{rap2} in order to
define finsheaves as finitary approximations of the sheaf
$\cont_{X}$ of continuous functions over the topological manifold
$X$. Indeed, the stalks of the latter, which host the germs of
continuous functions on $X$, were seen to arise as inductive
limits of the said finsheaves at the limit of infinite topological
refinement of the underlying open covers $\gauge_{i}$.

\item {\bf $\mathbf{fcqv}$-ADG-gravity.} Parenthetically, in closing this subsection,
we must note the significant physical import of fintriads in
ADG-gravity. So far, we have been able to formulate a {\em
manifestly background differential spacetime manifold independent
vacuum Einstein-Lorentzian gravity as a pure gauge
theory},\footnote{That is, a formulation of gravity solely in
terms of an algebraic $\struc$-connection field $\conn$.} with
finitistic, causal and quantum traits built into the theory from
the very beginning \cite{malrap1,malrap2,malrap3}. The high-point
in those investigations is that every fintriad $\triad_{i}$,
equipped with a finitary connection $\conn_{i}$ (:a finitary
instance of (\ref{eqx1})) and its associated curvature $\curv_{i}$
(:a finitary example of (\ref{eqx7})), is seen to support a
finitary version of the vacuum Einstein equations (\ref{eqx4}):

\begin{equation}\label{eqx5}
\ricci_{i}(\modl_{i})=0
\end{equation}

\noindent with geometric prequantization traits already attributed
to $\modl_{i}$ ({\it eg}, its local sections have been sheaf
cohomologically interpreted as quantum particle states of the
`field of quantum causality'---fittingly called `causons'
\cite{malrap2}).

Moreover, we have made thorough investigations on how ADG-gravity
can evade singularities of the most pathological
kind,\footnote{Like for instance when Rosinger's differential
algebras of generalized functions (:non-linear distributions),
hosting singularities everywhere densely in the background
topological space(time) $X$, are used as structure sheaves in the
theory (:`{\em spacetime foam differential triads}')
\cite{malros2,malros3}.} and their associated unphysical
infinities
\cite{malros1,malros2,mall3,malros3,mall9,mall7,mall11,mall10}
(especially \cite{malrap4}), with a special application to the
finitary-algebraic `resolution' of the inner Schwarzschild
singularity of the gravitational field of a point-particle
\cite{rap5}.

\end{itemize}

The $fcqv$-ADG-gravitational dynamics (\ref{eqx5}) may be thought
of as `taking place' within the category $\fintriad$, which in
turn may be regarded as a mathematical `universe' (`space') of
(dynamically) varying qausets. We shall return to comment more on
this in 4.3 and subsequent sections, after we show that
$\fintriad$ is actually a finitary example of an EGT---a fintopos.
The crux of the argument here is that as every sheaf (and {\it in
extenso} topos) of, say, sets, can be thought of as a
(mathematical) world of varying sets \cite{macmo}, so $\fintriad$
may be thought of as a universe of dynamically variable (qau)sets,
varying under the influence (action) of the
$fcqv$-ADG-gravitational field $\conn_{i}$. We thus first turn to
the topos-theoretic perspective on $\fintriad$ next.

\subsection{$\fintriad$ as a finitary example of an ET}

The title of the present subsection is one of the two main
mathematical results in the present paper---the other being that
$\fintriad$ is also a finitary instance of a GT-like structure, as
we shall show in the next section.

First, let us stress the following subtle point: $\fintriad$ is a
category of (fin)sheaves {\em not} over a fixed topological space
like the usual sheaf categories (:topoi) encountered in standard
topos theory \cite{macmo}, but over `{\em variable}' finitary
topological spaces (:fintoposets)---spaces that `vary' with
topological refinement (:coarse graining) and the associated
`degree $i\in I$ of topological resolution', as described
above.\footnote{This remark will prove to be important in the
sequel when we interpret $\fintriad$ as a finitary replacement of
the classical `continuum topos' $\sh^{0}(X)$, and in the last
section, where we shall remark on the possibility of relating our
scheme to Isham's `quantizing on a category' scenario.}

Now, the arrival at the result that $\fintriad$ is an ET (:a
cartesian closed category) is quite straightforward: one simply
has to juxtapose the properties of $\ctriad$, as they were
gathered from \cite{pap1,pap2,pap3,pap4,pap5} and presented at the
end of subsection 2.2, against the formal definitional properties
of an ET {\it \`a la} Lawvere and Tierney, as taken from
\cite{macmo} and synoptically laid out in appendix \ref{app1} at
the end. Then, one should bring forth that $\fintriad$ is a full
subcategory of $\ctriad$, enjoying all the latter's formal
properties. Thus, to recapitulate these properties, $\fintriad$ is
an ET because it:

\begin{itemize}

\item is closed under finite limits and colimits (:it is finitely
bicomplete); moreover, it is closed even `asymptotically' ({\it
ie}, under infinite topological refinement of the base $P_{i}$s
and their underlying $\gauge_{i}$s) in $\ctriad$, as we saw
earlier;

\item has finite (cartesian) products and coproducts (direct
sums);

\item has an exponential structure given, for any pair of
fintriads, by continuous maps $f_{ji}$ between the underlying
fintoposets and the fintriad morphisms that these maps lift to;
moreover, this structure `intertwines' canonically with the said
cartesian product as explained above; and finally,

\item has canonical subobjects. That is to say, it has a subobject classifier that
it inherits naturally from the archetypical topos $\sh(X)$ of
sheaves (of structureless sets, for example; or for instance, from
the topos $\sh^{0}(X)$ of sheaves $\cont_{X}$ of rings of
continuous $\mathbb{K}$-valued functions) over the original
topological manifold $X$. This will become more transparent in the
next section where we present $\fintriad$ alternatively as a GT.

\end{itemize}

\noindent An important question that arises from the exposition
above is the following: {\em what is the subobject classifier
$\Omega$ in $\fintriad$, and perhaps more importantly, what is its
physical interpretation?} For this, the reader will have to wait
for our ADG-based Grothendieck-type of perspective on Sorkin's
finitary scheme in the next section.

\section{An ADG-theoretic Grothendieck-type of perspective on Sorkin in $\fintriad$}

In this section we give an alternative topos-theoretic description
of the ET $\fintriad$. We present it as a finitary example of a
GT-like structure. In a way, a Grothendieck-type of perspective on
the finitary topology scenario of Sorkin \cite{sork0} is perhaps
more `canonical' and `natural' than the (more abstract) ET vantage
for viewing $\fintriad$ as a topos proper, because in Sorkin's
work, as we witnessed in the previous section, such notions as
{\em covering}, {\em sieve-topology} and its associated {\em
topological coarse graining} procedure, figure prominently in the
theory and they have well known, direct and generalized
correspondents in Grothendieck's celebrated work \cite{macmo}.

At the same time, by presenting $\fintriad$ as a type of GT will
enable us to see straightforwardly what the subobject classifier
$\sub$ is in it. With $\sub$ in hand, and by viewing it as a `{\em
generalized truth values object}' as it is its customary logical
interpretation in standard topos theory \cite{macmo}, we shall
then open paths for potentially exploring deep connections between
(spacetime) geometry ({\it eg}, topology) and (quantum) logic. In
fact, since $\fintriad$ carries {\em differential} geometric (not
just topological) information, and since in the past we have
successfully employed this structure to model $fcqv$-ADG-gravity,
the road will be open for investigating close relationships
between the differential geometric (`gravitational') structure of
the world, and its quantum logical traits. We shall explore two
such potential relationships with important physical
interpretation and implications in the epilogue. In the present
section however, we just present the topological refinement in
Sorkin's scheme by {\em differential geometric morphisms} in
$\fintriad$.

{\it Ex altis} viewed, and more from a technical (:mathematical)
vantage, invoking Grothendieck's categorical ideas in an
ADG-theoretic context appears to be only natural, since the
machinery of (abstract) {\em sheaf cohomology} is central in the
ADG-technology \cite{mall1,mall2,mall4}, while (abstract) {\em
sheaf cohomology} was originally the {\it raison d'\^{e}tre et de
faire} of Grothendieck's pioneering category-theoretic work in
general homological algebra. For it is no exaggeration to say that
Grothendieck's inspired vision, within the purely mathematical
`confines' of {\em algebraic} geometry, was to replace `space'
(:topology) by sheaf cohomology \cite{macmo}. Similarly, in
Mallios' ADG, now within the field of {\em differential} geometry
and with a strong inclination towards theoretical physics'
applications (especially in QG) \cite{mall1,mall4}, the ultimate
desire is to do away with the background geometrical (smooth)
spacetime (manifold) and the various (differential geometric)
anomalies (:singularities and related unphysical infinities) that
it carries \cite{malrap4,rap5}, and focus solely on the purely
algebraico-categorical (:sheaf-theoretically modelled) relations
between the `objects' that `live' on that surrogate
background---{\it ie}, the dynamical fields $\conn$ and the laws
(:differential equations) that they obey on their respective
carrier sheaves $\modl$, such as (\ref{eqx4})
\cite{malrap3,mall7,malrap4,mall11,mall10}.

\subsection{$\fintriad$ as a finitary instance of a GT}

As noted in the introduction and briefly described in 3.1 above,
Sorkin's main idea in \cite{sork0} was to `blow up' or `smear' the
points of the topological spacetime manifold $X$ by `fat' regions
(:open sets) $U$ about them belonging to locally finite open
covers $\gauge_{j}$ of $X$, and then to replace (:approximate) the
(locally) Euclidean $\cont$-topology of $X$, which is supposed to
be ``{\em carried by its points}'' \cite{sork0}, by
$T_{0}$-fintoposets $P_{j}$.

Such an enterprize has a rather natural correspondent and quite a
generalized description in category-theoretic terms. The latter
pertains to Grothendieck's celebrated work on generalized
topological spaces called {\em sites}, for the definition of which
{\em covering sieves} (associated with open covers in the usual
topological case), and a {\em Grothendieck topology} generated by
them play a central role \cite{macmo}.\footnote{See appendix
\ref{app2} at the end for the relevant (abstract) definitions.}
Thus, below we give a Grothendieck-type of description of Sorkin's
`finitarities', which will subsequently help us view $\fintriad$
as a GT of a finitary sort. In turn, in complete analogy to how
the $P_{j}$s in Sorkin's work were thought of as locally finite
approximations of the continuous topology of $X$, here the
EGT-like $\fintriad$ can be regarded as a finitary substitute of
the archetypical EGT $\sh^{0}(X)$---the topos of sheaves (of
rings) of continuous functions on the topological manifold $X$.
Thus, our research program of applying ADG-theoretic ideas to
finitary spacetime and gravity
\cite{malrap1,malrap2,malrap3,rap5,malrap4} is hereby reaching its
categorical (:topos-theoretic) climax.

So to begin with, let $X$ be the relatively compact
region\footnote{Recall that a topological space $X$ is said to be
relatively compact if every open cover of it admits a locally
finite refinement.} of a topological manifold $M$ that Sorkin
considered in \cite{sork0}, and $\gauge_{j}$ (a locally finite)
open cover for it, which also belongs to the inverse system
(:topological refinement net) $\invf :=\{ \gauge_{j}\}_{j\in
I}$.\footnote{In what follows, the reader should not confuse the
refinement index `$j$' used to label the fincovers in the
topological refinement net, with the subscript `$i$' labelling the
open sets in a particular covering. However, we shall use the same
symbol (:`$I$') to denote the index sets for both, hopefully
without causing any misunderstanding.} $X$ may be viewed as a {\em
poset category} $\poset(X)$, having for objects its open subsets
and for (monic) arrows between them subset-inclusions (:one arrow
for every pair of subsets, if they happen to be ordered by
set-theoretic inclusion):\footnote{In fact, $\poset(X)$ is more
than a poset, it is a {\em lattice}, but this will not concern us
in what follows.}

\begin{equation}\label{eq7}
U,V\subseteq X,~\mathrm{open}:~U\mapto V\Leftrightarrow U\subseteq
V
\end{equation}

\noindent Then, a {\em sieve} $S$ on $U$, $S(U)$, is an
$I$-indexed collection of open subsets of $U$ ($\{ V_{i\in
I}:~V_{i}\mapto U\}$) such that if $W\mapto V\in
S(U)\Longrightarrow W\in S(U)$. One moreover says that $S(U)$ {\em
covers} $U$ ({\it ie}, $S(U)$ is a {\em covering sieve} for the
object $U$ in $\poset(X)$), if $U\subseteq\bigcup_{i\in I}V_{i}\in
S(U)$. Arrow-wise, one says that $S(U)$ {\em covers the arrow}
$W\mapto U$ in $\poset(X)$ when $W\mapto\bigcup_{i}V_{i}$. With
the aid of the relevant abstract definitions in appendix
\ref{app2}, it is fairly straightforward to show for the concrete
category $\poset(X)$ that:

\medskip

\noindent {\bf Theorem:} The collection $\{(U,S(U))\}$, as $U$
runs through all the objects in $\poset(X)$, defines a
Grothendieck topology $J$ on $\poset(X)$.\footnote{This is just
exercise 1 on page 155 of \cite{macmo}.} Thence, the pair
$(\poset(X), J)$ is an example of a {\em site}---the poset
category $\poset{X}$ equipped with the Grothendieck topology $J$.

\medskip

Equivalently, calling to action the open covering $\gauge_{j}$ of
$X$ (and {\it in extenso} of $U$, since plainly,
$\bigcup_{i}V_{i}\longleftarrow U$), we can generate the following
{\em covering sieve} for $U$ based on $\gauge_{j}$:

\begin{equation}\label{eq8}
S_{j}\equiv S_{\gauge_{j}}(U)=\{ W\mapto U:~W\mapto
V_{i},~\mathrm{for~some}~V_{i}\in\gauge_{j}\}
\end{equation}

\noindent It follows then that, as $\gauge_{j}$ runs through the
inverse system (:refinement net) $\invf=\{\gauge_{j}\}_{j\in I}$,
a {\em basis} $\basis_{J}$ for the said Grothendieck topology $J$
on $\poset(X)$ is defined,\footnote{Again, see appendix \ref{app2}
for the relevant (abstract) definitions.} which also turns the
said poset category into the {\em site} $(\poset ,\basis_{J})$
\cite{macmo}.\footnote{As noted in appendix \ref{app2}, we hereby
do not distinguish between the site $(\poset(X),J)$ and
$(\poset(X),\basis_{J})$ generating it.}

\medskip

Now we have a good grasp of how Grothendieck-type of ideas can be
applied to $\poset(X)$ so as to promote it to a site. Thus, let us
turn to our category $\fintriad$ and see how it can qualify as a
finitary version of a GT-like structure.

{\it Prima facie}, and in view of the general and abstract ideas
presented in appendix \ref{app2}, we could maintain that a
`natural' two-step path one could follow in order to cast
$\fintriad$ as a finitary type of GT is the following:

\begin{itemize}

\item first head-on endow $\fintriad$ with some kind of {\em Grothendieck
topology} thus turn it into a {\em site}-like structure, as we did
for $\poset(X)$ above \cite{macmo};

\item then define {\em sheaves} over the resulting site and
collect them into a GT-like structure.

\end{itemize}

However, the alert reader could immediately counter-observe that:

\begin{itemize}

\item On a first sight, it appears to be hopeless to directly try and Grothendieck-topologize
the collection $\cov$ of {\em all} coverings $\gauge$ of the
continuum $X$ (or equivalently, the collection of all
subtopologies $\tau_{\gauge}$ of $X$ generated by them), since
that family is not even a set proper---{\it ie}, it is a {\em
class}. As a result, if one wished to view $\cov$ as some sort of
category, it would certainly {\em not} be small, unlike what the
usual Grothendieck categories are assumed to be.\footnote{See
appendix \ref{app2}. Similar reservations were expressed in
\cite{ish0}, where the poset of subtopologies of a continuum
appeared to be a class unmanageably large, hence unsuitable for
quantization. Thus Isham had to resort to {\em finite topologies}
(:topologies on a finite set of points) and the lattice of
subtopologies thereof for a plausible quantization scenario.}

\item Moreover, as noted earlier, $\fintriad$ is {\em
not} a category of sheaves over a {\em fixed} base topological
space, so that even if the latter was somehow
Grothendieck-topologized to a {\em site}, $\fintriad$ would still
{\em not} be a GT proper. Rather, the base spaces of the fintriads
in $\fintriad$ are `variable' entities, varying with the
topological refinement (:coarse graining) of the underlying
finitary coverings and their associated fintoposets.

\end{itemize}

Our way-out of this two-pronged impasse is based on the following
two observations:

\begin{enumerate}

\item First, in response to the first `dead-end' above, we note that the locally finite open covers
$\gauge_{i}$ of Sorkin are, categorically speaking, `good', `well
behaved' objects when it comes to defining some generalized kind
of `topology' on them and taking `limits' with respect to it. This
is due to the fact that the collection $\invf$ of all the finitary
coverings of $X$ comprise a so-called {\em cofinal} subset of the
class $\cov$ of all (proper) open covers of $X$
\cite{mall1,mall2}.

\item Second, and issuing from the point above, one could indeed use
the topological refinement partial order
($\gauge_{i}\preceq\gauge_{j}\Leftrightarrow
P_{j}\stackrel{f_{ji}}{\mapto}P_{i}$) on the net $\invf$ (and its
associated projective system $\inv$ of fintoposets)\footnote{One
can use $\invf$ and $\inv$ interchangeably, since one can transit
from the $\gauge_{i}$s in $\invf$ to the $P_{i}$s in $\inv$ by
Sorkin's `factorization algorithm' (\ref{eq2}).} so as to define
some kind of `{\em topological coarse graining sieve-topology}' on
it {\it \`a la} Grothendieck. Then indeed, Sorkin's inverse limit
`convergence' of the elements of $\invf$ (and their associated
$P_{i}$s) to $X$ at infinite topological refinement (\ref{eq5}),
can be accounted for on the grounds of that (abstract) topology.
In the process however, a new type of GT arises, which we call
`{\em a finitary  approximation topos}' (`fat')---one that may be
thought of as `approximating' the usual `continuum topos'
$\sh^{0}(X)$ of sheaves of (rings of) continuous functions over
the pointed $\cont$-manifold $X$, much in the same way that the
fintoposets $P_{i}$ were seen to approximate the continuum $X$ (or
equivalently, the finsheaves in \cite{rap2} were seen to
approximate the `continuum' sheaf $\cont_{X}$).\footnote{Note in
this respect that $\sh^{0}(X)$ may indeed be thought of as a GT if
we recall from above that $(\poset(X),J)$---or equivalently,
$(\poset(X),\basis_{J})$---is a site. At the same time,
$\sh^{0}(X)$ is a typical example of an ET as well \cite{macmo}.}

\end{enumerate}

\paragraph{$\fintriad$ as a `fat'-type of GT.} We {\em can} endow the poset category
$\invf$ with a Grothendieck-type of topology by introducing the
notion of {\em coarse graining finsieves}. Indirectly, these
played a significant role earlier, when we defined the basis
$\basis_{J}$ for the site $(\poset(X) ,J)$.

So, recall that the fincovers in $\invf$ are partially ordered by
refinement, $\gauge_{i}\preceq\gauge_{j}$, which is tantamount to
coarse graining continuous surjective maps (:arrows) between their
respective fintoposets, $f_{ji}:~P_{j}\mapto f_{i}$ (\ref{eq4}).
With respect to these arrows, and with appendix \ref{app2} as a
guide, we first define {\em coarse graining finsieves}
$S_{i}\equiv S_{\gauge_{i}}$ covering each and every object
(:fincover $\gauge_{i}$) in $\invf$,\footnote{$S_{i}:=\{
\gauge_{j}\in\invf:~\gauge_{j}\preceq \gauge_{i}\}$.} and from
these we also define {\em coarse graining finsieves} $S_{ji}$
covering each and every arrow $f_{ji}\in\invf$. With $S_{i}$ and
$S_{ji}$ in hand ($\forall\gauge_{i},f_{ji}\in\invf, ~i\in I$), we
then define a Gothendieck topology $J_{I}$ on $\invf$, thus
converting it to a site: $(\invf ,J_{I})$.\footnote{In fact, the
covering coarse graining finsieves defined above determine (object
and arrow-wise) a {\em basis} $\basis_{i}$ for $J_{I}$ (see
appendix \ref{app2}). As noted before, we think of $(\invf
,J_{I})$ and $(\invf ,\basis_{i})_{i\in I}$ as being equivalent
sites.} Parenthetically, as briefly alluded to earlier, the
central projective limit result of Sorkin about $\inv$
(\ref{eq5}), may now be literally understood as the {\em
`convergence' of the cofinal system $\invf$ of finitary coverings,
at the limit of their infinite topological refinement, to $X$
relative to the Grothendieck-type of topology $J_{I}$ (or the
Grothendieck basis $(\basis_{i})_{i\in I}$) imposed on it}.

To unveil the GT-like character of $\fintriad$, now that the net
$\invf$ of base spaces of its objects (:fintriads) has been
Grothendieck-topologized, is fairly straightforward. We simply
recall that $\fintriad$ is a category of finsheaves of incidence
algebras over Sorkin's fintoposets (:fintriads) deriving from the
$\gauge_{i}$s in the Grothendieck net $\invf$; hence, it is a
finitary example of a GT (:a category of sheaves over a site).
Moreover, because the inverse limit space of the $P_{i}$s is
effectively ({\it ie}, modulo Hausdorff reflection) homeomorphic
to $X$, and with our original regarding finsheaves as finitary
approximations of $\cont_{X}$ (for $X$ a topological manifold), we
may think of $\fintriad$ as a `fat' of (the EGT)
$\sh^{0}(X)$---the category of sheaves of (rings of) continuous
functions over the $\cont$-manifold $X$.

\paragraph{$\fintriad$ is coherent and localic.} At this point it is important to throw in this
presentation some technical remarks in order to emphasize that
$\fintriad$ is manifestly ({\it ie}, by construction) {\em
finitely generated}; hence, in a finitary sense, {\em coherent}
\cite{macmo}. Moreover, by the way finsheaves were defined in
\cite{rap2} ({\it ie}, as `skyscraper'-like, fat/coarse {\it
\'{e}tale} spaces over the coarse, blown-up `points' of $X$
corresponding to the minimal open sets/nerves covering them
relative to a locally finite cover $\gauge_{i}$ of
$X$),\footnote{Indeed, as noted before, in Sorkin's work the
points of $X$ were substituted by
$\stackrel{\gauge_{i}}{\sim}$-equivalence classes (and $X$ by the
corresponding $P_{i}$s).} $\fintriad$ {\em has enough points} and
it is {\em localic} \cite{macmo}. Indeed, its underlying locale
$\loc$\footnote{A {\em locale} is a complete distributive lattice,
otherwise known as a {\em Heyting algebra}. Locales are usually
thought of as `{\em pointless topological spaces}'. It is a
general result that every GT has an underlying locale
\cite{macmo}.} is just the lattice of open subsets of $X$, while
the points of $X$ are recovered (modulo Hausdorff reflection) at
the projective limit of infinite refinement of the base
$\gauge_{i}$s (or their associated $P_{i}$s) of the fintriads as
we go along the coarse graining Grothendieck-type of sieve
topology on $\fintriad$.\footnote{In the general case of an
abstract coherent topos, there is a celebrated result due to
Deligne stipulating roughly that {\em every coherent topos has
enough points and its underlying locale is a topological space
proper} \cite{macmo}.}

We can thus exploit the said `localicality' of $\fintriad$ in
order to find out what is its subobject classifier $\sub$. We do
this next.

\subsection{The subobject classifier in the EGT $\fintriad$}

That $\fintriad$ is localic points to a way towards its subobject
classifier. One may think of the base spaces of the fintriads in
$\fintriad$ as `{\em finitary locales}' (finlocales) since, as
noted earlier, they are the `pointless' subtopologies $\tau_{i}$
of $X$ generated by arbitrary unions of finite intersections of
the open sets in each locally finite open cover $\gauge_{i}$ of
$X$. We also noted that $\fintriad$ can be regarded as a `fat' of
$\sh^{0}(X)$.\footnote{In this respect there is an intended
metaphorical pun between the acronym `fat' (:finitary
approximation topos) and the epithet `fat'. Indeed, the
Grothendieck fintopos $\fintriad$ associated with Sorkin's
finitarities comes from substituting $X$'s points by {\em fat},
coarse open regions about them (and hence the pointed $X$ by the
pointless finlocales $\tau_{i}$).} The generic object in the
latter is $\cont_{X}$---the sheaf of continuous functions on the
pointed topological manifold $X$. When $X$ is
Grothendieck-topologized and turned into a site as described
earlier, $\cont_{(\poset ,J)}$ is a sheaf on a site and hence
$\sh^{0}(\poset ,J)$ {\em the} canonical example of an EGT
\cite{macmo}.

Now, a central result in topos theory is the following: {\em for
any sheaf on a site $X$, the lattice $\loc$ of all its subsheaves
is a complete Heyting algebra, a locale}.\footnote{See theorem on
page 146 in \cite{macmo}.} Thus, in our case we just take
$\cont_{X}$ (which is a generic object in $\sh^{0}(X)$) for the
said `sheaf-on-a-site', and the finsheaves (in the fintriads) for
its subsheaves (:subobjects). Plainly then, the subobject
classifier $\Omega$ in $\fintriad$ is

\begin{equation}\label{eq9}
\Omega(\fintriad)=\loc(\cont_{X})\equiv\loc(\cont_{(\poset(X)
,J)})
\end{equation}

\noindent hence $\fintriad$ is a localic topos, as anticipated
above. In terms of covering sieves in the standard case of a
topological space $X$ like ours (again, regarded as a poset
category $\poset(X)$, with objects its open subsets $U$), we
borrow verbatim from \cite{macmo}\footnote{Page 140.} that ``{\em
for sheaves on a topological space with the usual open cover
topology, the subobject classifier is the sheaf $\omg$ on $X$
defined by:} $\omg(U)=\{ V|~V~\mathrm{is~open~and}~V\subset
U\}$'', or in terms of covering (:principal) sieves
$S_{\downarrow}$ of lower sets for every $V$ as above ({\it ie},
$S_{\downarrow}(V):=\{ V^{'}|~V^{'}\subseteq V\}$), ``$\omg(U)=\{
S_{\downarrow}(V)\}$'' ({\it cf.} appendix \ref{app2}).

\paragraph{Interpretational matters: the semantic interplay between geometry and logic in a topos.}
One of the quintessential properties of a topos like $\sh^{0}(X)$
(for $X$ a $\cont$-manifold)---one that distinguishes it from the
topos $\set$ of `{\em constant sets}' \cite{macmo}---is that its
subobject classifier is a complete Heyting algebra, in
contradistinction to the Boolean topos $\set$ whose subobject
classifier $\Omega$ is the Boolean binary alternative
$\mathbf{2}=\{ 0,1\}$. This inclines one to `geometrically'
interpret the former topos, in contradistinction to $\set$, as
`{\em a generalized space of continuously variable sets, varying
continuously with respect to the background continuum $X$}'
\cite{lawvere,macmo}. In the same semantic vain, we may interpret
the variation of the objects living in $\fintriad$ (:qausets
\cite{malrap1,malrap2,malrap3,rap5,malrap4}) as {\em entities
varying with (topological) coarse graining}.

At the same time, every topos like $\sh^{0}(X)$ has not only a
{\em geometrical}, but also a {\em logical} interpretation due to
the non-Boolean character of its subobject classifier. Indeed, as
noted before, $\Omega$ {\em can also be regarded as a generalized
truth-values object}, the generalization being the transition from
the Boolean truth values $\Omega=\mathbf{2}=\{ 0,1\}=\{ \top
,\bot\}$ in $\set$, to a Heyting algebra-type of subobject
classifier like the one in $\fintriad$. This means that the
so-called `{\em internal language}' (or logic) that can be
associated with such topoi is (typed and) {\em intuitionistic}, in
contrast to the `classical', Boolean logic of the topos $\set$ of
sets \cite{macmo,lambek}.

In the last section we shall entertain the idea of exploring this
close connection between geometry and logic in our particular case
of interest (:$\fintriad$), and we shall briefly pursue its
physical implications and potential import to QG research.

However, for the time being, in the last paragraph of the present
section we would like to give an ADG-based topos-theoretic
presentation of topological refinement, which played a key role
above in viewing $\fintriad$ as a GT-like structure.

\paragraph{Finitary differential geometric morphisms: topological refinement
as a natural transformation from the differential geometric
standpoint of ADG.} Regarding the {\em differential} geometric
considerations that come hand in hand with ADG, since the
fintriads encode not only topological, but also differential
geometric structure, we may give a differential geometric flavor
to Sorkin's purely topological acts of refinement in $\invf$
and/or $\inv$.

We may recall from section 2 that, from a general topos-theoretic
vantage, a continuous map $f:~X\mapto Y$ between two topological
spaces gives rise to a pair $\geomorph_{f}=(f_{*},f^{*})$ of
covariant adjoint functors between the respective sheaf categories
(:topoi) $\sh_{X}$ and $\sh_{Y}$ on them, called push-out ({\it
alias}, direct image) and pull-back ({\it alias}, inverse image).
In topos-theoretic jargon, $\geomorph$ is known as a {\em
geometric morphism}. In our case of interest (:$\fintriad$), the
continuous surjection $f_{ji}:~P_{j}\mapto P_{i}$ (equivalently
regarded as the map $f_{ji}:~\tau_{j}\mapto\tau_{i}$)
corresponding to topological coarse graining (or equivalently, to
covering refinement $\gauge_{i}\preceq\gauge_{j}$), induces via
the Sorkin-Papatriantafillou scenario a pair
$\geomorph_{f_{ji}}\equiv\geomorph_{ji}=(f_{ji*},f_{ji}^{*})$ of
{\em fintriad morphisms} between the fintriads $\triad_{i}$ and
$\triad_{j}$. $\geomorph_{ji}$ by definition (of differential
triad morphisms) preserves the differential structure encoded in
the finsheaves (of incidence algebras) comprising the
corresponding fintriads, thus it may be called {\em finitary
differential geometric morphism}. Thus, $\fintriad$ may be
perceived as a category whose objects are $\triad_{i}$s and whose
arrows are $\geomorph_{ji}$s. The latter give a differential
geometric slant to Sorkin's purely topological acts of refinement.

Furthermore, since the sheaves defining the fintriads are
themselves functors\footnote{Every sheaf (of any structures, {\it
eg}, sets, groups, vector spaces, rings, modules {\it etc}.) on a
topological space may be identified with the (associated) {\em
sheafification functor} between the respective categories ({\it
eg}, from the category of topological spaces to that of groups)
that produces it \cite{mall1,macmo}.} the functors
$(f_{ji*},f_{ji}^{*})$ between the general sheaf categories (of
sets) $\sh_{\tau_{i}}\equiv \sh_{i}\ni\triad_{i}$ and
$\sh_{\tau_{j}}\equiv\sh_{j}\ni\triad_{j}$ may be thought of as
{\em natural transformations}, and hence $\fintriad$ as a type of
{\em functor category} \cite{macmo}. {\it In summa}, Sorkin's
topological refinement may be understood in terms of ADG as a kind
of {\em natural transformation of a differential geometric
character}---we may thus coin it `{\em differential geometric
refinement}'.

\subsection{Functoriality: general covariance is preserved under refinement}

Differential geometric refinement has a direct application and
physical interpretation in ADG-gravity. In
\cite{malrap3,rap5,malrap4} we saw how the Principle of General
Covariance (PGC) of GR can be expressed categorically in
ADG-theoretic terms as the $\struc$-{\em functoriality} of the
vacuum Einstein equations (\ref{eqx4}) or its finitary analogue
(\ref{eqx5}). This means that (\ref{eqx4}) is expressed via the
curvature $\ricci$, which is an $\struc$-morphism or
$\struc$-tensor (where $\otimes_{\struc}$ is the homological
tensor product functor with respect to $\struc$). The physical
significance of the $\struc$-functoriality of the ADG-theoretic
vacuum gravitational dynamics is that our choice of
field-measurements or field-coordinatizations encoded in $\struc$
({\it in toto}, our choice of $\struc$), does not affect the field
dynamics.\footnote{In turn, in \cite{malrap3,rap5} and especially
in \cite{malrap4}, this $\struc$-functoriality of the field
dynamics was taken to support the ADG Principle of Field Realism
(PFR): the connection field $\conn$, expressed and partaking into
the gravitational dynamics (\ref{eqx4}) via its curvature, is not
`perturbed' by our measurements/coordinatizations in $\struc$.}
More familiarly, the ADG-analogue of the
$\mathrm{Diff}(M)$-implemented PGC of the differential manifold
$M$ based GR, is $\aut_{\struc}\modl$---the principal (group)
sheaf of field automorphisms. Since $\modl$ is by definition
locally coordinatized (`Cartesianly analyzed') into
$\struc$,\footnote{Recall from section 2 that $\modl$ is defined
as a locally free $\struc$-module of finite rank: $\modl
:\stackrel{\mathrm{loc.}}{\simeq}\struc^{n}$.} $\aut_{\struc}\modl
|_{U\subset X}:=\modl
nd\modl(U)^{\bull}=M_{n}(\struc)^{^{\bull}}(U)\equiv\mathcal{G}\mathcal{L}(n,\struc)(U)\equiv
\mathrm{GL}(n,\struc(U))$, and the ADG-version of the PGC is
(locally) implemented via
$\mathcal{G}\mathcal{L}(n,\struc)(U)\equiv
\mathrm{GL}(n,\struc(U))$---the (local) `$\struc$-analogue' of the
usual $GL(4,\R)$ of GR standing for the group of general (local)
coordinates' transformations.

Now, in \cite{rap5,malrap4} it was observed that the said
generalized coordinates' $\struc$-independence
(:$\struc$-functoriality) of the ADG-gravitational field dynamics
has a rather natural categorical representation in terms of {\em
natural transformations} (pun intended). Diagrammatically, this
can be represented as follows:

\[
\bfig
\putsquare<1`1`1`1;900`900>(900,900)[\triad_{1}\ni\struc_{1}(:\modl_{1}\stackrel{\mathrm{loc.}}{\simeq}\struc_{1}^{n})`\ricci(\modl_{1})=0
`\triad_{2}\ni\struc_{2}(:\modl_{2}\stackrel{\mathrm{loc.}}{\simeq}\struc_{2}^{n})`\ricci(\modl_{2})=0;
\otimes_{\struc_{1}}`\natf_{\struc}`\natf_{\conn}`\otimes_{\struc_{2}}]
\efig
\]

\noindent and it reads that, changing (via the natural
transformation $\natf$) structure sheaves of algebras of
generalized arithmetics (coordinates) from an $\struc_{1}$ (and
its corresponding $\modl_{1}$) to another $\struc_{2}$ (and hence
$\modl_{2}$), the functorially, $\otimes_{\struc}$-expressed (via
the connection's curvature $\struc$-morphism) vacuum Einstein
equations remain `form invariant'.\footnote{Parenthetically, note
in the diagram above that $\natf$ has two indices: one for the
`object' (:structure sheaf $\struc$), and one for the `morphism'
(:connection $\conn$, or better, its $\struc$-morphism curvature
$\ricci$) in the respective differential triads $\triad_{1}$ and
$\triad_{2}$ supporting them. By definition, a natural
transformation between such sheaf (:functor) categories, is a
functor that preserves objects (:sheaves) and their morphisms
(:connections and their curvatures).}

The upshot here is that, the differential geometric refinement in
$\fintriad$ described above may be perceived as such a natural
transformation-type of map (:finitary differential geometric
morphism), as follows:

\[
\bfig
\putsquare<1`1`1`1;1000`1000>(1000,1000)[\triad_{j}\ni\struc_{j}(:\modl_{j}\stackrel{\mathrm{loc.}}{\simeq}\struc_{j}^{n})`\ricci_{j}(\modl_{j})=0
`\triad_{i}\ni\struc_{i}(:\modl_{i}\stackrel{\mathrm{loc.}}{\simeq}\struc_{i}^{n})`\ricci_{i}(\modl_{i})=0;
\otimes_{\struc_{j}}`\natf_{ji\struc}\equiv\geomorph_{ji\struc}`\natf_{ji\conn}\equiv\geomorph_{ji\conn}`\otimes_{\struc_{i}}]
\efig
\]

\noindent Thus, the ADG-version of the PGC of GR
(:$\struc$-functoriality) is preserved under such a not only
topological, but also differential geometric, refinement. It is
precisely this result that underlies the inverse system
$\inveinst$ of vacuum Einstein equations and its continuum
projective limit in the tower of inverse/direct systems of various
finitary ADG-structures in expression (150) of \cite{malrap3}
and/or (25) of \cite{rap5}. In the latter paper especially, it is
the projective limit of $\inveinst$ that was used to argue that
the vacuum Einstein equations hold over the inner Schwarzschild
singularity of the gravitational field of a point-particle both at
the finitary (`discrete') and at the classical continuum inverse
limit of infinite refinement (of the underlying base fintoposets
of the fintriads involved).

\section{Epilogue cum Speculation: Four Future QG Prospects}

In this rather lengthy concluding section we elaborate on the
following four promising future prospects. First, on how one might
further build on the EG-fintopos so as to incorporate `quantum
logical' ideas into our scheme. Then, we ponder on potential
affinities between our ADG-based finitary EGT $\fintriad$ and, (i)
Isham's recent `quantizing on a category' scenario, (ii)
Christensen-Crane's recent causite theory, and (iii)
Kock-Lawvere's SDG.

\subsection{Representation theory: associated Hilbert fintopos
$\hilbert$}

As noted earlier, by now it has been appreciated (primarily by
mathematicians!) that a topos can be regarded both as a
generalized space in a geometrical ({\it eg}, topological) sense,
as well as a generalized logical universe of variable set-like
entities. Thus, in a topos, `geometry' and `logic' are thought of
as being unified \cite{macmo}.

In our case, in view of this geometry-logic unification in a
topos, a future prospect for further developing the theory is to
relate the (differential) geometric (:`gravitational') information
encoded in the fintopos $\fintriad$, to the (internal) logic of an
`{\em associated Hilbert fintopos}' $\hilbert$. The latter may be
obtained from $\fintriad$ in three steps:

\begin{enumerate}

\item From \cite{aigner,stanley,zap0} first invoke finite dimensional (irreducible)
Hilbert space $\hil_{i}$ matrix representations for every
incidence algebra $\omg_{i}$ dwelling in the stalks of every
finsheaf $\Omg_{i}$.

\item Then, like the corresponding incidence algebras were stacked into the finsheaves $\Omg_{i}$, group the
$\hil_{i}$s into {\em associated} \cite{vas3,vas4}
(:representation) Hilbert finsheaves $\hilb_{i}$ (again over
Sorkin's fintoposets, which are subject to topological
refinement).

\item Finally, organize the $\hilb_{i}$s into the fintopos
$\hilbert$ as we did for the $\Omg_{i}$s in $\fintriad$, which may
be fittingly coined the {\em Hilbert fintopos associated to}
$\fintriad$.

\end{enumerate}

\noindent What one will have effectively obtained in the guise of
$\hilbert$ is a {\em coarse graining presheaf of Hilbert spaces}
(:a presheaf of Hilbert $D$-modules
\cite{hilbert,kato-1,kato0,kato2}) over the topological refinement
poset category $\inv$ (or $\invf$). To see this clearly, one must
recall from \cite{rapzap1,rapzap2,zap1} that the correspondence
$P_{i}\mapto\omg_{i}$ is a {\em contravariant functor} from the
poset category $\inv$ and the continuous (:monotone) maps
(:fintoposet morphisms) $f_{ji}$ between the $P_{i}$s, to a direct
(:inductive) system $\diromg$ of finitary incidence algebras and
surjective algebra homomorphisms $\omega_{ji}$ between them. Such
a contravariant functor may indeed be thought of as a {\em
presheaf} \cite{macmo}.\footnote{This remark will prove to be
useful in the next subsection.}

\paragraph{A `unified' perspective on geometrical and logical obstructions.} The
pair $(\fintriad ,\hilbert)$ of fintopoi may provide us with
strong clues on how to unify the `warped' (gravitational) geometry
and the `twisted' (quantum) logic in a topos-theoretic setting. In
this respect, the following analogy between the two topoi in the
pair above is quite suggestive:

\begin{itemize}

\item As we saw in \cite{malrap1,malrap2,malrap3}, the finsheaves
$\Omg_{i}$ in $\fintriad$ admit non-trivial (gravitational)
connections $\conn_{i}$, whose curvature $\curv_{i}(\conn_{i})$
measures some kind of {\em obstruction} preventing the following
sequence of generalized differentials (:connections)

\begin{equation}\label{eqz1}
\begin{array}{c}
\mathbf{\Omega^{0}}(\modl)\stackrel{\conn\equiv\conn^{0}}{\mapto}\mathbf{\Omega^{1}}(\modl)
\stackrel{\conn^{1}}{\mapto}
\mathbf{\Omega^{2}}(\modl)\stackrel{\conn^{2}}{\mapto}\mathbf{\Omega^{3}}(\modl)
\stackrel{\conn^{3}}{\mapto}\cdots\cr
\cdots\stackrel{\conn^{i-1}}{\mapto}\mathbf{\Omega^{i}}(\modl)\stackrel{\conn^{i}}{\mapto}\mathbf{\Omega^{i+1}}(\modl)
\stackrel{\conn^{i+1}}{\mapto}\cdots
\end{array}
\end{equation}

\noindent from being {\em exact}. This is in contrast to the usual
de Rham complex

\begin{equation}\label{eqz2}
\begin{array}{c}
\mathbf{0}(\equiv\mathbf{\Omega^{-2}})\stackrel{\imath\equiv
d^{-2}}{\mapto}\mathbf{K}(\equiv\mathbf{\Omega^{-1}})\stackrel{\epsilon\equiv
d^{-1}}{\mapto}\struc(\equiv\mathbf{\Omega^{0}})\cr
\stackrel{d^{0}\equiv\partial}{\mapto}\mathbf{\Omega^{1}}\stackrel{d^{1}\equiv\kd}{\mapto}
\mathbf{\Omega^{2}}\stackrel{d^{2}}{\mapto}\cdots\mathbf{\Omega^{n}}
\stackrel{d^{n}}{\mapto}\cdots
\end{array}
\end{equation}

\noindent which is exact in our theory (:finitary de Rham theorem)
\cite{malrap2}. In other words, the curvature $\curv$ of the
connection $\conn$ measures the departure of the latter from
flatness, as opposed to $\partial$ which is flat.\footnote{For
example, section-wise in the relevant finsheaves:
$(\conn_{i}^{j+1}\circ\conn_{i}^{j})(s\otimes
t)=t\wedge\curv_{i}(s)$, with $s\in\Gamma(U,\Omg_{i})$,
$t\in\Gamma(U,\mathbf{\Omega^{j}})$ and $U$ open in $X$. Thus,
$\curv_{i}(\conn_{i})$ represents not only the measure of the
departure from differentiating flatly, but also the deviation from
setting up an (exact) cohomology sequence based on
$\conn_{i}$---altogether, a measure of the departure of
$\conn_{i}$ from (the) nilpotence (of $\partial_{i}$).}
Equivalently, in topos-theoretic parlance, the finsheaves in
$\fintriad$ do not have {\em global elements} (:sections)
\cite{macmo}.\footnote{Page 164.} In $\fintriad$, absence of
global sections of its curved finsheaves is captured by the
non-existence of arrows from the terminal object $\mathbf{1}$ in
the topos to the said finsheaves. In ADG-theoretic terms
\cite{mall1,mall2,malrap2}, section-wise the obstruction
(:departure from exactness) of the $\conn$-complex above due to
$\curv(\conn)$, may be expressed via the non-triviality of the
`{\em global section functor}' and of the complex

\begin{equation}\label{eqz3}
\begin{array}{c}
\Gamma_{X}(\mathcal{S}^{^\centerdot}):~~\Gamma_{X}(\mathbf{0})\mapto\Gamma_{X}(\mathbf{\mathcal{S}^{0}})
\stackrel{\Gamma_{X}(d^{0})}{\mapto}\Gamma_{X}(\mathbf{\mathcal{S}^{1}})
\stackrel{\Gamma_{X}(d^{1})}{\mapto}\cdots\cr
\cdots\stackrel{\Gamma_{X}(d^{n-1})}{\mapto}\Gamma_{X}(\mathbf{\mathcal{S}^{n-1}})
\stackrel{\Gamma_{X}(d^{n})}{\mapto}\Gamma_{X}(\mathbf{\mathcal{S}^{n}})\not{\!\!\!\!\mapto}\cdots
0
\end{array}
\end{equation}

\noindent that it defines \cite{mall1,mall2,malrap2}. Again, this
is a fancy way of saying that the relevant vector (fin)sheaves
($\mathcal{S}_{i}^{j}\equiv\Omg_{i}^{j},~j\in\mathbb{Z}_{+}$) do
not admit global sections due to the non-triviality of $\conn$.
All this has been physically interpreted as {\em absence of global
`inertial' frames} (:`inertial observers') in ADG-fingravity
\cite{malrap1}.

\item In a similar vain, but from a quantum logical standpoint, the associated Hilbert differential module finsheaves
$\hilb_{i}$ do {\em not} admit global sections (:`valuation
states')\footnote{For $\mathrm{dim}\hil_{i} >2$.} in view of the
Kochen-Specker theorem in standard quantum logic
\cite{buttish1,buttish2,buttish3,buttish4}. This is due to the
well known fact that there are maximal Boolean subalgebras
(:frames) of the the quantum lattice $\mathcal{L}_{i}(\hil_{i})$
that are generated by mutually incompatible (:complementary,
noncommuting) elements of $\mathcal{B}(\hil_{i})$---the
non-abelian $C^{*}$-algebra of bounded operators on
$\mathcal{L}_{i}$ (whose hermitian elements are normally taken to
represent quantum observables). The result is that certain
presheaves (of sets) over the coarse graining poset of Boolean
subalgebras of $\mathcal{L}_{i}(\hil_{i})$ do not admit global
sections. Logically, this is interpreted as saying that there are
no global (Boolean) truth values in quantum logic, but only local
ones ({\it ie}, `localized' at every maximal Boolean subalgebra or
frame of $\mathcal{L}_{i}(\hil_{i})$); moreover, the resulting
`truth values' space (:object) $\Omega$ in the corresponding
presheaf topos ceases to be Boolean (:$\Omega=\mathbf{2}$) and
becomes intuitionistic (:a Heyting algebra). In this sense,
quantum logic is contextual (:`Boolean subalgebra localized') and
`neorealist' (:not Boolean like the classical logic of $\set$, but
intuitionistic). Accordingly, in the aforesaid tetralogy of Isham
{\it et al.}, it has been explicitly anticipated that {\em there
must be a characteristic form that}, like $\curv_{i}(\conn_{i})$
above, {\em effectuates the said obstruction to assigning values
to physical quantities globally over $\mathcal{L}_{i}$}.

\item Thus, what behooves us in the future is to look for what one
might call a `{\em quantum logical curvature}' characteristic form
$\corv$ which measures {\em both} the (differential) geometrical
obstruction in $\fintriad$ to assigning global (inertial) frames
at its finsheaves $\Omg_{i}$, {\em and} the quantum logical
obstruction to assigning global (Boolean) frames to their
associated Hilbert finsheaves $\hilb_{i}$. This effectively means
that one could attempt to bring together the intuitionistic
(differential) geometric coarse graining in $\fintriad$, with the
also intuitionistic (quantum) logical coarse graining in
$\hilbert$.

\end{itemize}

One might wish to approach this issue of logico-geometrical
obstructions in a unified algebraic way. For instance, one could
observe that both the differential geometric obstruction (in GR)
and the quantum logical obstruction (in QM) above are due to some
non-commutativity in the basic `variables' involved, in the
following sense:

\begin{itemize}

\item the differential geometric obstruction, represented by the
curvature characteristic form, is due to the non-commutativity of
covariant derivations (:connections); while,

\item the quantum logical obstruction is ultimately due to the
existence of non-commuting (complementary) quantum observables
such as position (:$x$) and momentum (:$\partial_{x}$).

\end{itemize}

\noindent Parenthetically, we note in this line of thought that
for quite some time now the idea has been aired that the
`macroscopic' non-commutativity of covariant derivatives in the
curved spacetime continuum of GR is due to a more fundamental
`microscopic' quantum non-commutativity in a `discrete', dynamical
quantum logical (:`quantal') substratum underlying
it.\footnote{David Finkelstein in private e-mail correspondence
(2000).} For instance, in \cite{sel1,sel2,sel3,sel4} one witnesses
how the gravitational curvature form of a spin-Lorentzian
(:$SL(2,\com)$-valued) connection arises `spontaneously' (as a
coherent state condensate) from a Schwinger-type of dynamical
variational principle of basic bivalent spinorial quantum-time
atoms (:`chronons') teeming the said reticular and quantal
substratum coined the `quantum net' \cite{df2,df6,df5,df1}. In
this model we can quote Finkelstein from the prologue of
\cite{df1} maintaining that ``{\em logics come from dynamics}''.
For similar ideas, but in a topos-theoretic setting, see
\cite{rap0}.

On the other hand, in ADG-gravity there is no such fundamental
distinction between a (classical) continuum and a (quantal)
discretum spacetime. All there exists and is of import in the
theory are the algebraic (dynamical) relations between the
ADG-fields $(\modl ,\conn)$ themselves, without dependence on an
external (to those fields) surrogate background space(time), be it
`discrete/quantal' or `continuous/classical'
\cite{malrap1,malrap2,malrap3,malrap4,rap5}. Thus in our
ADG-framework, if we were to investigate deeper into the
possibility that some sort of quantum commutation relations are
ultimately responsible for the aforementioned obstructions, we
should better do it `sheaf cohomologically'---{\it ie}, in a
purely algebraic manner that pays respect to the fact that ADG is
not concerned at all with the geometrical structure of a
background spacetime, but with the algebraic relations of the
`geometrical objects' that live on that physically fiducial base.
The latter are nothing else than the connection fields $\conn$ and
the sections of the relevant sheaves $\modl$ that they act on,
while at the same time sheaf cohomology is {\em the} technical
(:algebraic) machinery that ADG employs from the very beginning of
the {\em aufbau} of the theory \cite{mall1,mall2,mall4,malrap2}.

We thus follow our noses into the realm of the ADG-perspective on
geometric (pre)quantization and second quantization
\cite{mall1,mall2,mall5,mall6,malrap2,mall4} in order to track the
said obstructions in $(\fintriad ,\hilbert)$ down to algebraic,
{\em sheaf cohomological commutation relations}. What we have in
mind is to propose some sheaf cohomological commutation relations
between certain characteristic forms that uniquely characterize
the finsheaves (and the connections acting on them) in the
fintopos $\fintriad$; while, by the functoriality of geometric
pre- and second quantization {\it \`a la} ADG
\cite{mall3,mall4,mall5,mall6,malrap2}, to transfer these
characteristic forms and their algebraic commutation relations to
their associated Hilbert finsheaves in $\hilbert$.

The following discussion on how we might go about and set up the
envisaged sheaf cohomological commutation relations is tentative
and largely heuristic.

One can begin by recalling some basic (`axiomatic') assumptions in
ADG-field theory \cite{mall1,mall2,malrap2,malrap3,malrap4}:

\begin{itemize}

\item The fields ({\it viz.} connections) $\conn$ exist `out
there' independently of us---the observers or `measurers' of them
(Principle of Field Realism in \cite{malrap3,malrap4}). Recall
from section 2 that in ADG-field theory, by a {\em field} we refer
to the pair $(\modl ,\conn)$. The connection $\conn$ is the `{\em
proper}' part of the field, while the vector sheaf $\modl$ is its
{\em representation} ({\it alias}, carrier or action) space.

\item Various collections $\gauge_{i}$ of covering open subsets of
the base topological space $X$ are the {\em systems of local open
gauges}.

\item Our measurements (of the fields) take values in the
structure sheaf $\struc$ of generalized arithmetics (coordinates
or coefficients) that {\em we} choose in the first place, with
$\struc(U)$ (for a $U$ in some $\gauge_{i}$ chosen) the {\em local
coordinate gauges} that we set up for ({\it ie}, to measure) the
fields.

\item From a geometric pre- and second quantization vantage
\cite{mall2,mall5,mall6,malrap2,malrap3,mall4}, our
field-measurements correspond to {\em local (particle)
coordinatizations of the fields}. They are the ADG-analogues of
`{\em particle position measurements}' of the fields. Local
position (particle) states are represented by local sections of
the representation (:associated) sheaves
$\modl$,\footnote{Furthermore, with respect to the spin-statistics
connection, local boson states are represented by local sections
of {\em line} sheaves (:vector sheaves of rank $1$), while
fermions by local sections of vector sheaves of rank greater than
$1$ \cite{mall1,mall2,mall5,mall6,malrap2,malrap3}.} which in turn
are by definition locally ($\struc_{U}$-) isomorphic to
$\struc^{n}$ ({\it ie}, $\modl(U)\equiv\modl
|_{U}\simeq(\struc(U))^{n}\equiv(\struc |_{U})^{n}$). Accordingly,
given a local gauge $U$ in a chosen gauge system $\gauge_{i}$, the
collection $e^{U}=\{(U;e_{1},\ldots ,e_{n})\}$ of local sections
of $\modl$ on $U$ is called a {\em local frame} (or local gauge
basis) of $\modl$. Any section $s\in\Gamma(U,\modl)\equiv\modl(U)$
can be written as a linear combination of the $e_{\alpha}$s above,
with coefficients in $\struc(U)$.

\item As noted before, $\modl$ is the carrier or action space of
the connection. $\conn$ acts on the local particle
(coordinate-position) states ({\it ie}, the local sections) of
$\modl$ and changes them. Thus, the (flat) sheaf morphisms
$\partial$, and {\it in extenso} the curved ones $\conn$, are the
generalized (abstract), ADG-theoretic analogues of {\em momenta}.

\end{itemize}

\noindent With these abstract semantic correspondences:

\begin{equation}\label{eqa1}
\begin{array}{c}
\mathrm{1.~abstract~position/particle~states}\mapto\mathrm{local~sections~of}~\modl\cr
\mathrm{2.~abstract~momentum/field~states}\mapto\mathrm{local~expression~of}~\conn
\end{array}
\end{equation}

\noindent we are in a position to identify certain characteristic
forms that could engage into the envisaged (local) quantum
commutation relations (relative to a chosen family $\gauge$ of
local gauges):

\begin{enumerate}

\item Concerning the abstract analogue of `particle/position'
(:the part $\modl$ of the ADG-field pair $(\modl ,\conn)$), we
might consider the so-called coordinate $1$-cocycle
$\phi_{\alpha\beta}\in\aut\modl=GL(n,\struc(U_{\alpha\beta}))=\mathcal{G}\mathcal{L}(n,\struc)(U_{\alpha\beta})$
(for $U_{\alpha\beta}=U_{\alpha}\cap
U_{\beta};~U_{\alpha},U_{\beta}\in\gauge$), which completely
characterizes (and classifies) sheaf cohomologically the vector
sheaves $\modl$ \cite{mall1,mall2,malrap2,mall4}.\footnote{Indeed,
we read from \cite{mall2} for example in connection with the
Picard cohomological classification of the vector sheaves involved
in ADG, that ``{\em any vector sheaf $\modl$ on $X$ is uniquely
determined (up to an $\struc$-isomorphism) by a coordinate
$1$-cocycle, say, $(g_{\alpha\beta})\in
Z^{1}(\gauge,\mathcal{G}\mathcal{L}(n,\struc))$, associated with
any local frame $\gauge$ of $\modl$}''.} What we have here is an
instance of the age-old Kleinian dictum that local states
(:`geometry') of $\modl$---{\it ie}, the (local) sections that
comprise it,\footnote{And recall {\em the} epitome of sheaf
theory, namely, that {\em a sheaf is its (local) sections}. That
is, the entire sheaf space $\modl$ can be (re)constructed (by
means of restriction and collation) from its (local) sections.
Local information (:sections) is glued together to yield the
`total sheaf space'.} are {\em how} they transform (here, under
changes of local gauge $\phi_{\alpha\beta}$).\footnote{Another way
to express this Kleinian viewpoint, $\modl$ is the associated
(:representation) sheaf of the `symmetry' group sheaf
$\aut\modl=\mathcal{G}\mathcal{L}(n,\struc)$ of its
self-transmutations. Equivalently, the particle states (:local
sections of $\modl$) of the field carry a representation of the
symmetry group of field automorphisms. Here, the epithet
`symmetry' pertains to the fact that $\aut\modl$ is the symmetry
group sheaf of vacuum Einstein ADG-gravity (\ref{eqx4}),
implementing our abstract version of the PGC of GR.}

\item Concerning the abstract analogue of `field/momentum' (:the
part $\conn$ of the ADG-field pair $(\modl ,\conn)$), we might
consider the so-called gauge potential $\aconn$ of the connection
$\conn$, which completely determines $\conn$
locally.\footnote{Indeed, we read from \cite{mall2} that ``{\em
$\conn$ is determined (locally) uniquely by $\aconn$}''. Recall
also that $\conn$ locally splits as $\partial +\aconn$
\cite{mall1,mall2,malrap3}.} With respect to $\gauge$,
$\aconn_{ij}$ is (locally) a $0$-cochain of (local) $n\times n$
matrices with entries from (:local sections in) $\Omg^{1}(U)$
($U\in\gauge$; $n$ is the rank of $\modl$). In other words, for a
given system (frame) of local gauges
$\gauge=(U_{\alpha})_{\alpha\in I}$, $\aconn^{(\alpha)}_{ij}\in
C^{0}(\gauge , M_{n}(\Omega)=C^{0}(\gauge,\Omega^{1}(\modl
nd\modl)$---{\it ie}, $\aconn$ is (locally) an endomorphism
(:$\modl nd\modl$) valued `$1$-form'.\footnote{In this respect,
one may recall that in the usual theory (CDG of smooth manifolds),
the gauge potential is (locally) a Lie algebra-valued $1$-form.}

\end{enumerate}

In line with the above, we may thus posit (locally) the following
abstract (pre)quantum commutation relations between the
generalized (:abstract) `{\em position characteristic form}'
$\phi_{\alpha\beta}$ and the generalized (:abstract) `{\em
momentum characteristic form}' $\aconn_{ij}$:\footnote{With
indices omitted.}

\begin{equation}\label{eqa2}
[\phi |_{U} ,\aconn |_{U}]\propto\corv\in\modl nd\modl(U)
\end{equation}

\noindent which make sense ({\it ie}, they are well defined),
since $\phi$ (locally) takes values in
$\aut\modl(U)=\mathcal{G}\mathcal{L}(n,\struc)(U)$, while $\aconn$
in $\modl nd\modl$, which both allow for (the definition of a
Lie-type of) a product like the commutator.

What behooves us now is to give a physical interpretation to the
commutator above according to the ADG-field semantics. Loosely,
(\ref{eqa2}) is the relativistic and covariant ADG-gravitational
analogue of the Heisenberg uncertainty relations between the
position and momentum `observables' of a (non-relativistic)
quantum mechanical particle (or, {\it in extenso}, of a
relativistic quantum field). One should highlight here a couple of
things concerning (\ref{eqa2}):

\begin{itemize}

\item First a mathematical observation: the functoriality between $\fintriad$ (gravity; differential geometry)
and $\hilbert$ (quantum theory; quantum logic) carries the
characteristic forms and their uncertainty relations from the
former to the latter.

\item Second, a physical observation: the `{\em self-quantumness}' of the ADG-field $(\modl
,\conn)$. As it has been stressed many times in previous work
\cite{malrap3,rap5,malrap4}, the ADG-field $(\modl ,\conn)$ is a
dynamically autonomous, `already quantum' and in need of no formal
process of quantization. The autonomy pertains to the fact that
there is no background geometrical spacetime (continuum or
discretum) interpretation of the purely algebraic, dynamical
notion of ADG-field (:ADG-gravity is a genuinely background
independent theory). Moreover, the field is `self-quantum' (or
`self-quantized') as its two constituent parts---$\modl$ and
$\conn$---engage into the quantum commutation relations
(\ref{eqa2}), while its background spacetime independence entails
that in our scheme {\em quantization of gravity is not dependent
on or does not entail quantization of spacetime itself}.

\item Since in ADG-gravity there is no background spacetime (continuous or discrete) interpretation,
while all is referred to the algebraic (dynamical) relations in
sheaf space, {\em there is no spacetime scale dependence of the
ADG-expressed law of vacuum Einstein gravity} (\ref{eqx4}), {\em
or of the commutator} (\ref{eqa2}). Recalling from the
introduction our brief remarks about the `conspiracy' of the
equivalence principle of GR and the uncertainty principle of QM,
which apparently prohibits the infinite localization of the
gravitational field past the so-called Planck space-time
length-duration without creating a black hole; by contrast, {\em
in ADG-gravity the Planck space-time is not thought of as a
fundamental `obstruction'---an unavoidable regularization cut-off
scale---to infinite localization beyond which the classical
continuum spacetime gives way to a quantal discretum one}. As
noted in \cite{malrap3,malrap4,rap5}, the vacuum Einstein
equations hold both at the classical continuum (\ref{eqx4}) and at
the quantal discontinuum level (\ref{eqx5}), and they are not
thought of as breaking down below Planck scale.

\item If any `noncommutativity' is involved in ADG-gravity (say, {\it \`a la} Connes \cite{connes}), it is encoded in
$\aut\modl$ (`field foam' \cite{malros2}), or anyway, in $\modl
nd\modl$ where $\phi$ and $\aconn$ take their values. That is, in
our scenario, if any kind of `noncommutativity' is involved, it
pertains to the dynamical self-transmutations of the field
(:$\conn$) and its `inherent' quantum particle states (:local
sections of $\modl$) \cite{malrap4}.\footnote{Recall that in
ADG-gravity the PGC of GR is modelled after $\aut\modl$, while
from a geometric prequantization viewpoint, the local quantum
particle states of the ADG-gravitation field (:the local sections
of $\modl$) are precisely the ones that are `shuffled around' by
$\aut\modl$---the states on which $\conn$ acts to dynamically
change.}

\item Since the sheaf cohomological quantum commutation relations
(\ref{eqa2}) are preserved by topological (or differential
geometric) refinement and are carried intact to the `classical
continuum limit' \cite{rapzap1,rapzap2}, we may interpret the
usual (differential geometric) curvature obstruction (in
$\fintriad$) as some kind of `macroscopic quantum effect' (coming
from $\hilbert$), like Finkelstein has intuited for a long while
now.\footnote{See footnote 46 above.}

\end{itemize}

\subsection{Potential links with Isham's quantizing on a category}

A future project of great interest is to relate our
fintopos-theoretic labors on ADG-fingravity above with Isham's
recent `{\em Quantizing on a Category}' (QC) general mathematical
scheme \cite{ish5,ish6,ish7,ish8}.

On quite general grounds, the algebraico-categorical QC is closely
akin to ADG both conceptually and technically, having affine basic
motivations and aims. For example, QC's main goal is to quantize
systems with configuration (or history) spaces consisting of
`points' having internal (algebraic) structure. The main
motivation behind QC is the grave failure of applying the
conventional quantization concepts and techniques to `systems'
({\it eg}, causets or spacetime topologies) whose configuration
(or general history) spaces are far from being
structureless-pointed differential (:smooth) manifolds. Isham's
approach hinges on two innovations: first it regards the relevant
entities as objects in a category, and then it views the
categorical morphisms as abstract analogues of momentum
(derivation maps) in the usual (manifold based) theories. As it is
also the case with ADG, although this approach includes the
standard manifold based quantization techniques, it goes much
further by making possible the quantization of systems whose
`state' spaces are not pointed-structureless smooth continua.

As hinted to above, there appear to be close ties between QC and
ADG-gravity---ties which ought to be looked at closer in the
future. {\em Prima facie}, both schemes concentrate on evading the
(pathological) point-like base differential manifold---be it the
configuration space of some classical or quantum physical system,
or the background spacetime arena of classical or quantum (field)
physics---and they both employ `pointless', categorico-algebraic
methods. Both focus on an abstract (categorical) representation of
the notion of derivative or derivation: in QC Isham abstracts from
the usual continuum based notion of vector field (derivation), to
arrive at the categorical notion of arrow field which is a map
that respects the internal structure of the categorical objects
one wishes to focus on ({\it eg}, topological spaces or causets);
while in our work, the notion of derivative is abstracted and
generalized to that of an algebraic connection, defined
categorically as a sheaf morphism, on a sheaf of suitably
algebraized structures ({\it eg}, causets or finitary topological
spaces and the incidence algebras thereof).

A key idea that could potentially link QC with our fintopos
$\fintriad$ for ADG-fingravity (and with ADG in general) is that
in the former, as a result of a formal process of quantization
developed there, a {\em presheaf of Hilbert spaces} (of variable
dimensionality) arises as a `induced representation space' of the
so-called `{\em category quantization monoid of arrow-fields}'
defined by the arrow-semigroup of the base category $C$ that one
chooses to work with.\footnote{Roughly, as briefly mentioned
above, the objects of $C$ in Isham's theory represent generalized
(:abstract) `configuration states', while the
transformation-arrows (:morphisms) between them, analogues of
momentum (:derivation) maps.} The crux of the argument here is
that this presheaf is similar to the coarse `graining Hilbert
presheaf' that the associated (:representation) Hilbert fintopos
$\hilbert$ was seen to determine above. This similarity motivates
us to wish to apply Isham's QC technology to our particular case
of interest in which the base category is $\fintriad$ and the
arrows between them the coarse graining fintriad geometric
morphisms, taking also into account the internal structure of the
finsheaves involved. In this respect, perhaps also the sheaf
cohomological quantization algebra envisaged above can be related
to the category (monoid) quantization algebras engaged in QC. All
in all, we will have in hand a particular application of Isham's
QC scenario to the case of our ADG-perspective on Sorkin's
fintoposets, their incidence algebras, and the finsheaves
(:fintriads) thereof in $\fintriad$.

\subsection{Potential links with the Christensen-Crane causites}

As noted in the introduction, recently there has been a
Grothendieck categorical-type of approach to quantum spacetime
geometry and to non-perturbative Lorentzian QG called {\em causal
site} (:causite) theory \cite{crane}. Causite theory bears a close
resemblance in both motivation and technical (:categorical) means
employed with our $fcqv$-ADG-gravity---in particular, with the
present topos-theoretic version of the latter. Here are some
common features:

\begin{itemize}

\item Both employ general homological algebra (:category-theoretic) ideas and techniques.
Causite theory may be perceived as a `categorification' and
quantization of causet theory, while our scheme may be understood
as the `sheafification' and (pre)quantization of causets.

\item In both approaches, simplicial ideas and techniques are
central. In causite theory there are two main structures, both of
which are modelled after partial orders: the topological and the
causal. The relevant categorical structures of interest are
bisimplicial $2$-categories. As a result, the Grothendieck-type of
topos structure envisaged to be associated with (pre)sheaves (of
Hilbert spaces) over causites is a $2$-topos (:bitopos). In our
approach on the other hand, the causal and topological structures
of the world are supposed to be physically indistinguishable,
hence they `collapse' into a single (simplicial) partial order.
This subsumes our main position that {\em the physical topology is
the causal topology} \cite{malrap1}. As a result, the fintopos
$\fintriad$-organization of the finsheaves of qausets over
Sorkin's finsimplicial complexes (:fintoposets) accomplished
herein is a Grothendieck-type of `unitopos', not a bitopos.

\item We read from \cite{crane}: ``{\em A very important feature of the topology of causal sites
is that they have a tangent $2$-bundle, which is analogous to the
tangent bundle of a manifold}''. In the purely algebraic
ADG-gravity, we are not interested in such a geometrical
interpretation and conceptual imagery (:base spacetime, tangent
space, tangent bundle {\it etc}). Presumably, one would like to
have a tangent bundle-like structure in one's theory in order to
identify its sections with `derivation maps', thus have in one's
hands not only {\em topological}, but also {\em differential}
structure. Having differential geometric structure on causites,
then one would like ``{\em to impose Einstein's equation}'' (as a
{\em differential} equation proper!) ``{\em on a causal site
purely intrinsically}''. Moreover, in \cite{crane} it is observed
that, in general, ``{\em doing sheaf theory over such generalized
spaces (:sites) is an important part of modern mathematics}.'' In
ADG-fingravity, we do most (if not all) of the above {\em entirely
algebraically}, {\it a fortiori} without any geometrical
commitment to a background `{\em space(time)}'---be it discrete or
continuous.

\item Last but not least, both approaches purport to be inherently finitistic and hence {\it ab initio}
free from singularities and other unphysical infinities. The
ultimate aim (or hope!) of causite theory is to ``{\em lead to a
description of quantum physics free from ultraviolet divergences,
by eliminating the underlying point set continuum}'' \cite{crane}.
So is ADG-gravity's \cite{malrap1,malrap2,malrap3,malrap4,rap5}.

\end{itemize}

\noindent It would certainly be worthwhile to investigate closer
the conceptual and technical affinities between causite theory and
$fcqv$-ADG-gravity.\footnote{Louis Crane in private
e-correspondence.}

\subsection{Potential links with Kock-Lawvere's SDG}

From a purely mathematical perspective, but with applications to
QG also in mind, it would be particularly interesting to see how
can one carry out under the prism of our EG-fintopos $\fintriad$
the basic finitary, ADG-theoretic constructions {\em internally}
in the said fintopos by using the `esoteric', intuitionistic-type
of language (:logic) of this topos exposed in this paper. For
example, by stepping into the constructive world of the topos, one
could bypass the `problem' (because
$\struc$-functoriality-violating) of defining {\em derivations} in
ADG.\footnote{Chris Mulvey in private e-correspondence.} {\it En
passant}, as briefly alluded to in footnote 4 and in the previous
subsection, the reader will have already noticed that no notion of
`{\em tangent vector field}' is involved in ADG---{\it ie}, no
maps in $Der:~\struc\mapto\struc$ are defined, as in the classical
geometrical manifold based theory (CDG). Loosely, this can be
justified by the fact that in the purely algebraic ADG the
(classical) geometrical notion of `tangent space' to the
(arbitrary) simply topological base space $X$ involved in the
theory, has essentially no meaning, but more importantly, {\em no
physical significance}, since $X$ itself plays no role in the
(gravitational) equations defined as differential equations proper
via the derivation-free ADG-machinery.

Even more importantly for bringing together ADG and SDG, and
having delimited the topos-theoretic (:intuitionistic-logical)
background underlying both ADG and SDG, one can then compare the
notion of {\em connection}---arguably, {\em the} key concept that
actually qualifies either theory as being a {\em differential}
geometry proper---as this concept appears in a categorical guise
in both theories
\cite{kock1,kock,laven,mall1,mall2,mall4,vas1,vas2,vas4}. For the
definition of the synthetic differential (:connection) $\partial$,
the intuitionistic internal logic of the `formal smooth topoi'
involved plays a central role, dating back to Grothendieck's
stressing the importance of `rings with nilpotent elements' in the
context of {\em algebraic} geometry. At the same time, for the
definition of $\partial$ as a sheaf morphism in ADG, no serious
use has so far been made of the intuitionistic internal logic of
the sheaf categories in which the relevant sheaves live. One
should thus wait for an explicit construction of those sheaves
from within ({\it ie}, by using the internal language of) the
relevant topoi. This is a formidable task well worth exploring;
for, applications' wise, recall again for instance from the
previous subsection the following words from \cite{crane}:

\begin{quotation}
\noindent ``{\small ...As yet, we do not know how to impose
Einstein's equation on a causal site purely intrinsically...}''
\end{quotation}

\noindent On the other hand, we certainly know how to (and we
actually do!) impose (\ref{eqx4})---or its finitary version
(\ref{eqx5})---from within the objects (:finsheaves) comprising
$\fintriad$, but without having made actual use of the latter's
internal logic.

To wrap up the present paper, we would like to recall from the
conclusion of \cite{stachel1}\footnote{With the original citation
being \cite{fraenkel}.} the following `prophetic' exchange between
Abraham Fr\"{a}nkel and Albert Einstein:\footnote{This author
wishes to thank John Stachel for timely communicating
\cite{stachel1} to him. This quotation, with an extensive
discussion on how it anticipates our application of ADG-ideas to
finitistic QG, as well as the former's potential links with SDG,
can be found in \cite{malrap4}.}

\begin{quotation}
\noindent ``{\small ...In December 1951 I had the privilege of
talking to Professor Einstein and describing the recent
controversies between the (neo-)intuitionists and their
`formalistic' and `logicistic' antagonists; {\em I pointed out
that the first attitude would mean a kind of atomistic theory of
functions, comparable to the atomistic structure of matter and
energy. Einstein showed a lively interest in the subject and
pointed out that to the physicist such a theory would seem by far
preferable to the classical theory of continuity. I objected by
stressing the main difficulty, namely, the fact that the
procedures of mathematical analysis, e.g., of differential
equations, are based on the assumption of mathematical continuity,
while a modification sufficient to cover an
intuitionistic-discrete medium cannot easily be imagined. Einstein
did not share this pessimism and urged mathematicians to try to
develop suitable new methods not based on continuity}\footnote{Our
emphasis.}...}''
\end{quotation}

\noindent A modern-day version of the words above, which also
highlights the close affinity between sheaf and topos theory
\cite{macmo} {\it vis-\`a-vis} QT and QG, is due to
Selesnick:\footnote{Steve Selesnick (private correspondence).}

\begin{quotation}
\noindent ``{\small ...One of the primary technical hurdles which
must be overcome by any theory that purports to account, on the
basis of microscopic quantum principles, for macroscopic effects
(such as the large-scale structure of what appears to us as
space-time, {\it ie}, gravity) is the handling of the transition
from `localness' to `globalness'. In the `classical' world this
kind of maneuver has been traditionally effected either
measure-theoretically---by evaluating largely mythical integrals,
for instance---or geometrically, through the use of sheaf theory,
which, surprisingly, has a close relation to topos theory. The
failure of integration methods in traditional approaches to
quantum gravity may be ascribed in large measure to the
inappropriateness of maintaining a manifold---a `classical'
object---as a model for space-time, while performing quantum
operations everywhere else. If we give up this classical manifold
and replace it by a quantal structure, then the already
considerable problem of mediating between local and global (or
micro and macro) is compounded with problems arising from the
appearance of subtle effects like quantum entanglement, and more
generally by the problems arising from the non-objective nature of
quantum `reality'...}''
\end{quotation}

Most of the discussion in this long epilogue has been highly
speculative, largely heuristic, tentative and incomplete, thus it
certainly requires further elaboration and scrutiny. However, we
feel that further advancing our theory on those four QG research
fronts in the near future is well worth the effort.

\section*{Acknowledgments}

The author is indebted to Chris Isham for once mentioning to him
the possibility of adopting and developing a generalized,
Grothendieck-type of perspective on Sorkin's work \cite{sork0}, as
well as for his unceasing moral and material support over the past
half-decade. He also thanks Tasos Mallios for orienting, guiding
and advising him about selecting and working out what may prove to
be of importance to QG research from the plethora of promising
mathematical physics ideas that ADG is pregnant to---in this paper
in particular, about the potentially close ties between ADG and
topos theory {\it vis-\`a-vis} QG. Numerous exchanges with
Georgios Tsovilis on the potential topos-theoretic development of
ADG-gravity are also acknowledged. Maria Papatriantafillou is also
gratefully acknowledged for e-mailing to this author her papers on
the categorical aspects of ADG \cite{pap1,pap2,pap3,pap4}, from
which some beautiful {\it LaTeX} commutative diagrams were
obtained. Finally, he wishes to acknowledge financial support from
the European Commission in the form of a European Reintegration
Grant (ERG-CT-505432) held at the University of Athens, Greece.

\appendix

\section{The Definition of an Abstract Elementary Topos}\label{app1}

This appendix is used in 3.2 to show that $\fintriad$ is a
finitary example of an ET in the sense of Lawvere and Tierney
\cite{macmo}. To recall briefly this formal and abstract
mathematical structure,\footnote{For more technical details, the
reader is referred to \cite{macmo}.} a small category\footnote{A
category is said to be small if the families of objects and arrows
that constitute it are proper sets---{\it ie}, not classes
\cite{macmo}.} $\cat$ is said to be an ET if it has the following
properties:

\begin{itemize}

\item $\cat$ is closed under finite limits. Equivalently, $\cat$
is said to be {\em finitely complete}. As noted earlier,
categorical limits are also known as projective (inverse) limits,
thus a topos $\cat$ is defined to be closed under projective
limits.

\item $\cat$ is {\em cartesian}. That is, for any two objects $A$ and
$B$ in $\cat$, one can form the object $A\times B$---their
cartesian product. All such finite products are supposed to be
`computable' in $\cat$ (:$\cat$ is closed under finite cartesian
products).

\item $\cat$ has an {\em exponential structure}. This essentially means
that for any two objects $A,B\in\cat$, one can form the object
$B^{A}$ consisting of all arrows (in $\cat$) from $A$ to $B$. As
noted earlier, the set $B^{A}$ is usually designated by
$\Hom(A,B)$ (:`hom-sets of arrows'), and it is supposed to
effectuate the following canonical isomorphisms for an arbitrary
object $C$ in $\cat$ relative to the cartesian product structure:
$\Hom(C\times A, B)\simeq\Hom(C,B^{A})$ (or equivalently:
$B^{C\times A}\simeq (B^{A})^{C}$).

\item $\cat$ has a {\em subobject classifier} object $\omg$. This means
that for any object $A$ in $\cat$, its subobjects (write
$\mathrm{sub}(A)$) canonically correspond to arrows from it to
$\omg$: $\mathrm{sub}(A)\simeq\Hom(A,\omg)\equiv\omg^{A}$.

\end{itemize}

\noindent A couple of secondary, `corollary' properties of a topos
$\cat$ are:

\begin{itemize}

\item $\cat$ is also {\em finitely cocomplete}. That is, $\cat$ is also
closed under finite inductive (direct) limits. Thus {\it in toto},
a topos $\cat$ is defined to be {\em finitely bicomplete}
(co-complete or co-closed).

\item $\cat$ has a preferred object $\mathbf{1}$, called the
{\em terminal object}, over which all the other objects in $\cat$
are `fibered'. That is, for any $A\in\cat$, there is a unique
morphism $A\mapto\mathbf{1}$.

\item Dually, $\cat$ also possesses a so-called {\em initial object}
$\mathbf{0}$ which is `included' in each and every object of
$\cat$; write: $\mathbf{0}\mapto A,~(\forall A\in \cat)$.

\item Finally, again dually to the fact that a topos $\cat$ has (finite)
products, it also has (finite) coproducts.\footnote{For example,
in the category $\set$ of sets---the archetypical example of a
topos that other topoi aim at generalizing---the coproduct is the
disjoint union (or direct sum) of sets and it is usually denoted
by $\coprod$ (or $\bigoplus$). On the other hand, in the category
of (commutative) rings, or of $\mathbb{K}$-algebras, or even of
sheaves of such algebraic objects, the coproduct is the usual
tensor product $\otimes_{\mathbb{K}}$ (while the product remains
the cartesian product, as in the universe $\set$ of structureless
sets).}

\end{itemize}

\noindent Due to its possessing (i) finite cartesian products,
(ii) exponentials, and (iii) a terminal object, an ET $\cat$ is
said to be a {\em cartesian closed category}, an equivalent
denomination \cite{macmo}. Let it be noted here that the primary
definitional axioms for an ET above are not minimal. Indeed, a
small category $\cat$ need only possess finite limits, a subobject
classifier $\Omega$, as well as a so-called power object
$PB=\Omega^{A}$ (for every object $A\in\cat$), in order to qualify
as an ET proper. Then, the rest of the properties outlined above
can be derived from these three basic ones \cite{macmo}.

\section{The Definition of an Abstract Grothendieck Topos}\label{app2}

This appendix is used in 4.1 to show that $\fintriad$ is a GT
\cite{macmo}. To recall briefly this formal and abstract
mathematical structure,\footnote{Again, for more technical
details, the reader can refer to \cite{macmo}.} a small category
$\cat$ is said to be a GT if the following two conditions are met:

\begin{itemize}

\item There is a base category $\base$ endowed with a so-called {\em
Grothendieck topology} on its arrows. $\base$, thus topologized,
is said to be a {\em site}; and

\item Relative to $\base$, $\cat$ is a {\em sheaf
category}---{\it ie}, it is a category of sheaves over the site
$\base$.

\end{itemize}

\noindent Let us elaborate a bit further on these two defining
features of an abstract GT.

\paragraph{Grothendieck topologies: sites.} There are two (equivalent) definitions of a
Grothendieck topology on a category $\base$, which we borrow from
\cite{macmo}. Both use the notion of a {\em sieve}---in
particular, of so-called {\em covering sieves}. {\it Prima facie},
the use of covering sieves in defining a Grothendieck topology is
tailor-cut for $\fintriad$, which follows Sorkin's tracks in
\cite{sork0}, since we saw in 3.1 that the notions of {\em open
coverings} and {\em sieve-topologies} generated by them play a
central role in Sorkin's fintoposet scheme.

Thus, for an object $A$ in a category $\base$, a {\em sieve} $S$
on $A$ (write $S(A)$) is a set of arrows (:morphisms) $f:~*\mapto
A$ in $\base$\footnote{$*$ stands for an arbitrary object in
$\base$, which happens to be the domain of an arrow $f\in S(A)$.}
such that for all arrows $g\in\base$ with
$\mathrm{dom}(f)=\mathrm{ran}(g)$,\footnote{Where `$\mathrm{dom}$'
and `$\mathrm{ran}$' denote the `{\em domain}' and `{\em range}'
maps on the arrows of $\base$, respectively. That is, for
$\base\ni h:~B\mapto C$, $\mathrm{dom}(h)=B$ and
$\mathrm{ran}(h)=C$.}

$$f\in S(A)\Longrightarrow f\circ g\equiv fg \in S(A)$$

\noindent That is, $S$ is a {\em right ideal} in $\base$, when the
latter is viewed as an associative arrow-semigroup with respect to
morphism-multiplication (:arrow concatenation).

Parenthetically, in the case of a topological space
$X$,\footnote{In which we are interested in 4.1 in connection with
Sorkin's `finitarities' in \cite{sork0} and ours in $\fintriad$.}
regarded as a {\em poset category} $\poset(X)$ of its open subsets
$U\subseteq X$ and having as ({\em monic}) morphisms between them
open subset-inclusions ({\it ie},
$\forall~\mathrm{open}~U,V\subseteq X:~V\mapto U\Leftrightarrow
V\subseteq U$), a sieve on $U$ is a {\em poset ideal} (with
`$\subseteq$' the relevant partial order).

From the definition of a sieve above, it follows that if $S(A)$ is
a sieve on $A$ in $\base$, and $g:~B\mapto A$ any arrow with
$\mathrm{ran}(g)=A$, then the collection

$$g^{*}(S)=\{ \base\ni h:~\mathrm{ran}(h)=B,~gh\in S\}$$

\noindent is also a sieve on $B$ called the pull-back (sieve) of
(the sieve) $S$ along (the arrow) $g$.

Having defined sieves, an abstract kind of topology $J$---the
so-called {\em Grothendieck topology}---can be defined on a
general category $\base$ in terms of them. Thus, $J$ is an
assignment to every object $A$ in $\base$ of a family $J(A)$ of
sieves on $A$, satisfying the following three
properties:\footnote{The following is the `{\em object-form}'
definition of a Grothendieck topology \cite{macmo}. Its equivalent
`{\em arrow-form}' follows shortly.}

\begin{itemize}

\item {\bf Maximality:} the {\em maximal} sieve $m(A)=\{
f:~\mathrm{ran}(f)=A\}$ belongs to $J(A)$;

\item {\bf Stability:} if $S\in J(A)$, then $g^{*}(S)$ belongs to
$J(B)$ for any arrow $g$ as above;

\item {\bf Transitivity:} if $S\in J(A)$ and $T(A)$ is any sieve
on $A$ such that $\forall g$ as above, $g^{*}(T)\in J(B)$, then
$T\in J(A)$.

\end{itemize}

\noindent We say that $S$ {\em covers} $A$ (or that $S$ is a {\em
covering sieve} for $A$ relative to $J$ on $\base$), when it
belongs to $J(A)$. Also, we say that a sieve $S$ {\em covers the
arrow} $g:~B\mapto A$ above, if $g^{*}(S)\in J(B)$.

With these two `covering' definitions, and by identifying the
objects of $\base$ by their identity arrows $i_{A}:~A\mapto A$
($\forall A\in\base$), the three defining properties of a
Grothendieck topology on $\base$ above can be recast in `{\em
arrow-form}' as follows \cite{macmo}:

\begin{itemize}

\item {\bf Maximality:} if $S$ is a sieve on $A$ and $f\in S$,
then $S$ covers $f$;

\item {\bf Stability:} if $S$ covers an arrow $f:~B\mapto A$, it
also covers $g\circ f$, $\forall g:~C\mapto B$; and,

\item {\bf Transitivity:} if $S$ covers the arrow $f$ above, and
$T$ is a sieve on $A$ covering all the arrows in $S$, then $T$
covers $f$.

\end{itemize}

\noindent Finally, an instrumental notion (used in 4.1) is that of
a {\em basis} $\basis_{J}$ or {\em generating set of morphisms}
for a (covering sieve in a) Grothendieck topology $J$ on a general
category $\base$ with pullbacks. Following \cite{macmo},
$\basis_{J}$ is an assignment to every object $A\in\base$ of a
collection $\basis_{J}(A):=\{ f:~\mathrm{ran}(f)=A\}$ of arrows in
$\base$ with range $A$, enjoying the following properties:

\begin{itemize}

\item {\bf Iso-Maximality:} every isomorphism in $\base$, with range $A$, belongs to
$\basis_{J}(A)$;

\item {\bf Stability:} for a family $F=\{ f_{i}:~B_{i}\mapto A~(i\in I)\}$ in $\basis_{J}(A)$,
and any morphism $g:~C\mapto A$, the family of pullbacks $\{
f^{*}_{i}:~B_{i}\times_{A}C\mapto C\}$ along each $f_{i}$ belongs
to $\basis_{J}(C)$; and,

\item {\bf Transitivity:} for $F$ as above, and for each $i\in I$ one has another
family of arrows $G_{j}=\{ g_{ij}:~C_{ij}\mapto B_{i}~(j\in
I_{i})\}$ in $\basis_{J}(B_{i})$, the family $F\circ G:=\{
f_{i}\circ g_{ij}:~C_{ij}\mapto A~(i\in I,j\in I_{i})\}$ also
belongs to $\basis_{J}(A)$.

\end{itemize}

A category $\base$ equipped with a Grothendieck topology $J$ as
defined above is called a {\em site}. A site is usually symbolized
by the pair $(\base ,J)$. If instead of $J$ one has prescribed a
basis $\basis_{J}$ on $\base$, by slightly abusing terminology,
the pair $(\base ,\basis_{J})$ can still be called a
site---namely, it is the site generated by the {\em covering
families} of arrows in $\basis_{J}(A)$ ($\forall A\in\base$).

{\it In summa}, a site represents a generalized topological space
on which (abstract) sheaves can be defined. Indeed, as noted in
the main text, Grothendieck invented sites in order to develop
generalized {\em sheaf cohomology} theories thus be able to tackle
various problems in algebraic geometry \cite{macmo}.

\paragraph{Sheaves on a site: GT.} With a site $(\base ,J)$ in hand,
an abstract GT is defined to be a category $\cat$ of sheaves over
a base site. One writes symbolically, $\cat:=\sh(\base ,J)$.

\medskip

\noindent $\bullet$ It is a general fact that every GT is an ET
\cite{macmo}.\footnote{Page 143.}

\end{document}

%% file: diagram.tex
\makeatletter

\def\diagram{\m@th\leftwidth=\z@ \rightwidth=\z@ \topheight=\z@
\botheight=\z@ \setbox\@picbox\hbox\bgroup}

\def\enddiagram{\egroup\wd\@picbox\rightwidth\unitlength
\ht\@picbox\topheight\unitlength \dp\@picbox\botheight\unitlength
\hskip\leftwidth\unitlength\box\@picbox}

\def\bfig{\begin{diagram}}
\def\efig{\end{diagram}}
\newcount\wideness \newcount\leftwidth \newcount\rightwidth
\newcount\highness \newcount\topheight \newcount\botheight

\def\ratchet#1#2{\ifnum#1<#2 \global #1=#2 \fi}

\def\putbox(#1,#2)#3{%
\horsize{\wideness}{#3} \divide\wideness by 2 {\advance\wideness
by #1 \ratchet{\rightwidth}{\wideness}} {\advance\wideness by -#1
\ratchet{\leftwidth}{\wideness}} \vertsize{\highness}{#3}
\divide\highness by 2 {\advance\highness by #2
\ratchet{\topheight}{\highness}} {\advance\highness by -#2
\ratchet{\botheight}{\highness}} \put(#1,#2){\makebox(0,0){$#3$}}}

\def\putlbox(#1,#2)#3{%
\horsize{\wideness}{#3} {\advance\wideness by #1
\ratchet{\rightwidth}{\wideness}} {\ratchet{\leftwidth}{-#1}}
\vertsize{\highness}{#3} \divide\highness by 2 {\advance\highness
by #2 \ratchet{\topheight}{\highness}} {\advance\highness by -#2
\ratchet{\botheight}{\highness}}
\put(#1,#2){\makebox(0,0)[l]{$#3$}}}

\def\putrbox(#1,#2)#3{%
\horsize{\wideness}{#3} {\ratchet{\rightwidth}{#1}}
{\advance\wideness by -#1 \ratchet{\leftwidth}{\wideness}}
\vertsize{\highness}{#3} \divide\highness by 2 {\advance\highness
by #2 \ratchet{\topheight}{\highness}} {\advance\highness by -#2
\ratchet{\botheight}{\highness}}
\put(#1,#2){\makebox(0,0)[r]{$#3$}}}

\def\adjust[#1]{} 

\newcount \coefa
\newcount \coefb
\newcount \coefc
\newcount\tempcounta
\newcount\tempcountb
\newcount\tempcountc
\newcount\tempcountd
\newcount\xext
\newcount\yext
\newcount\xoff
\newcount\yoff
\newcount\gap%
\newcount\arrowtypea
\newcount\arrowtypeb
\newcount\arrowtypec
\newcount\arrowtyped
\newcount\arrowtypee
\newcount\height
\newcount\width
\newcount\xpos
\newcount\ypos
\newcount\run
\newcount\rise
\newcount\arrowlength
\newcount\halflength
\newcount\arrowtype
\newdimen\tempdimen
\newdimen\xlen
\newdimen\ylen
\newsavebox{\tempboxa}%
\newsavebox{\tempboxb}%
\newsavebox{\tempboxc}%

\newdimen\w@dth

\def\setw@dth#1#2{\setbox\z@\hbox{\m@th$#1$}\w@dth=\wd\z@
\setbox\@ne\hbox{\m@th$#2$}\ifnum\w@dth<\wd\@ne \w@dth=\wd\@ne \fi
\advance\w@dth by 1.2em}

\def\t@^#1_#2{\allowbreak\def\n@one{#1}\def\n@two{#2}\mathrel
{\setw@dth{#1}{#2} \mathop{\hbox to
\w@dth{\rightarrowfill}}\limits \ifx\n@one\empty\else
^{\box\z@}\fi \ifx\n@two\empty\else _{\box\@ne}\fi}}
\def\t@@^#1{\@ifnextchar_{\t@^{#1}}{\t@^{#1}_{}}}
\def\to{\@ifnextchar^{\t@@}{\t@@^{}}}

\def\t@left^#1_#2{\def\n@one{#1}\def\n@two{#2}\mathrel{\setw@dth{#1}{#2}
\mathop{\hbox to \w@dth{\leftarrowfill}}\limits
\ifx\n@one\empty\else ^{\box\z@}\fi \ifx\n@two\empty\else
_{\box\@ne}\fi}}
\def\t@@left^#1{\@ifnextchar_{\t@left^{#1}}{\t@left^{#1}_{}}}
\def\toleft{\@ifnextchar^{\t@@left}{\t@@left^{}}}

\def\two@^#1_#2{\allowbreak
\def\n@one{#1}\def\n@two{#2}\mathrel{\setw@dth{#1}{#2}
\mathop{\vcenter{\lineskip\z@\baselineskip\z@
                 \hbox to \w@dth{\rightarrowfill}%
                 \hbox to \w@dth{\rightarrowfill}}%
       }\limits
\ifx\n@one\empty\else ^{\box\z@}\fi \ifx\n@two\empty\else
_{\box\@ne}\fi}}
\def\tw@@^#1{\@ifnextchar _{\two@^{#1}}{\two@^{#1}_{}}}
\def\two{\@ifnextchar ^{\tw@@}{\tw@@^{}}}

\def\tofr@^#1_#2{\def\n@one{#1}\def\n@two{#2}\mathrel{\setw@dth{#1}{#2}
\mathop{\vcenter{\hbox to \w@dth{\rightarrowfill}\kern-1.7ex
                 \hbox to \w@dth{\leftarrowfill}}%
       }\limits
\ifx\n@one\empty\else ^{\box\z@}\fi \ifx\n@two\empty\else
_{\box\@ne}\fi}}
\def\t@fr@^#1{\@ifnextchar_ {\tofr@^{#1}}{\tofr@^{#1}_{}}}
\def\tofro{\@ifnextchar^ {\t@fr@}{\t@fr@^{}}}

\def\mon{\mathop{\m@th\hbox to
      14.6\P@{\lasyb\char'51\hskip-2.1\P@$\arrext$\hss
$\mathord\rightarrow$}}\limits} 
\def\leftmono{\mathrel{\m@th\hbox to
14.6\P@{$\mathord\leftarrow$\hss$\arrext$\hskip-2.1\P@\lasyb\char'50%
}}\limits} 
\mathchardef\arrext="0200       

\def\settypes(#1,#2,#3){\arrowtypea#1 \arrowtypeb#2 \arrowtypec#3}
\def\settoheight#1#2{\setbox\@tempboxa\hbox{#2}#1\ht\@tempboxa\relax}%
\def\settodepth#1#2{\setbox\@tempboxa\hbox{#2}#1\dp\@tempboxa\relax}%
\def\settokens`#1`#2`#3`#4`{%
     \def\tokena{#1}\def\tokenb{#2}\def\tokenc{#3}\def\tokend{#4}}
\def\setsqparms[#1`#2`#3`#4;#5`#6]{%
\arrowtypea #1 \arrowtypeb #2 \arrowtypec #3 \arrowtyped #4 \width
#5 \height #6 }
\def\setpos(#1,#2){\xpos=#1 \ypos#2}

\def\settriparms[#1`#2`#3;#4]{\settripairparms[#1`#2`#3`1`1;#4]}%

\def\settripairparms[#1`#2`#3`#4`#5;#6]{%
\arrowtypea #1 \arrowtypeb #2 \arrowtypec #3 \arrowtyped #4
\arrowtypee #5 \width #6 \height #6 }

\def\resetparms{\settripairparms[1`1`1`1`1;500]\width 500}

\resetparms

\def\mvector(#1,#2)#3{
\put(0,0){\vector(#1,#2){#3}}%
\put(0,0){\vector(#1,#2){26}}%
}
\def\evector(#1,#2)#3{{
\arrowlength #3
\put(0,0){\vector(#1,#2){\arrowlength}}%
\advance \arrowlength by-30
\put(0,0){\vector(#1,#2){\arrowlength}}%
}}

\def\horsize#1#2{%
\settowidth{\tempdimen}{$#2$}%
#1=\tempdimen \divide #1 by\unitlength }

\def\vertsize#1#2{%
\settoheight{\tempdimen}{$#2$}%
#1=\tempdimen
\settodepth{\tempdimen}{$#2$}%
\advance #1 by\tempdimen \divide #1 by\unitlength }

\def\putvector(#1,#2)(#3,#4)#5#6{{%
\ifnum3<\arrowtype \putdashvector(#1,#2)(#3,#4)#5\arrowtype \else
\ifnum\arrowtype<-3 \putdashvector(#1,#2)(#3,#4)#5\arrowtype \else
\xpos=#1 \ypos=#2 \run=#3 \rise=#4 \arrowlength=#5 \ifnum
\arrowtype<0
    \ifnum \run=0
        \advance \ypos by-\arrowlength
    \else
        \tempcounta \arrowlength
        \multiply \tempcounta by\rise
        \divide \tempcounta by\run
        \ifnum\run>0
            \advance \xpos by\arrowlength
            \advance \ypos by\tempcounta
        \else
            \advance \xpos by-\arrowlength
            \advance \ypos by-\tempcounta
        \fi
    \fi
    \multiply \arrowtype by-1
    \multiply \rise by-1
    \multiply \run by-1
\fi \ifcase \arrowtype
\or \put(\xpos,\ypos){\vector(\run,\rise){\arrowlength}}%
\or \put(\xpos,\ypos){\mvector(\run,\rise)\arrowlength}%
\or \put(\xpos,\ypos){\evector(\run,\rise){\arrowlength}}%
\fi\fi\fi }}

\def\putsplitvector(#1,#2)#3#4{
\xpos #1 \ypos #2 \arrowtype #4 \halflength #3 \arrowlength #3
\gap 140 \advance \halflength by-\gap \divide \halflength by2
\ifnum\arrowtype>0
   \ifcase \arrowtype
   \or \put(\xpos,\ypos){\line(0,-1){\halflength}}%
       \advance\ypos by-\halflength
       \advance\ypos by-\gap
       \put(\xpos,\ypos){\vector(0,-1){\halflength}}%
   \or \put(\xpos,\ypos){\line(0,-1)\halflength}%
       \put(\xpos,\ypos){\vector(0,-1)3}%
       \advance\ypos by-\halflength
       \advance\ypos by-\gap
       \put(\xpos,\ypos){\vector(0,-1){\halflength}}%
   \or \put(\xpos,\ypos){\line(0,-1)\halflength}%
       \advance\ypos by-\halflength
       \advance\ypos by-\gap
       \put(\xpos,\ypos){\evector(0,-1){\halflength}}%
   \fi
\else \arrowtype=-\arrowtype
   \ifcase\arrowtype
   \or \advance \ypos by-\arrowlength
       \put(\xpos,\ypos){\line(0,1){\halflength}}%
       \advance\ypos by\halflength
       \advance\ypos by\gap
       \put(\xpos,\ypos){\vector(0,1){\halflength}}%
   \or \advance \ypos by-\arrowlength
       \put(\xpos,\ypos){\line(0,1)\halflength}%
       \put(\xpos,\ypos){\vector(0,1)3}%
       \advance\ypos by\halflength
       \advance\ypos by\gap
       \put(\xpos,\ypos){\vector(0,1){\halflength}}%
   \or \advance \ypos by-\arrowlength
       \put(\xpos,\ypos){\line(0,1)\halflength}%
       \advance\ypos by\halflength
       \advance\ypos by\gap
       \put(\xpos,\ypos){\evector(0,1){\halflength}}%
   \fi
\fi }

\def\putmorphism(#1)(#2,#3)[#4`#5`#6]#7#8#9{{%
\run #2 \rise #3 \ifnum\rise=0
  \puthmorphism(#1)[#4`#5`#6]{#7}{#8}#9%
\else\ifnum\run=0
  \putvmorphism(#1)[#4`#5`#6]{#7}{#8}#9%
\else
\setpos(#1)%
\arrowlength #7 \arrowtype #8 \ifnum\run=0 \else\ifnum\rise=0
\else \ifnum\run>0
    \coefa=1
\else
   \coefa=-1
\fi \ifnum\arrowtype>0
   \coefb=0
   \coefc=-1
\else
   \coefb=\coefa
   \coefc=1
   \arrowtype=-\arrowtype
\fi \width=2 \multiply \width by\run \divide \width by\rise \ifnum
\width<0  \width=-\width\fi \advance\width by60 \if l#9
\width=-\width\fi
\putbox(\xpos,\ypos){#4}
{\multiply \coefa by\arrowlength
\advance\xpos by\coefa \multiply \coefa by\rise \divide \coefa
by\run \advance \ypos by\coefa
\putbox(\xpos,\ypos){#5} }%
{\multiply \coefa by\arrowlength
\divide \coefa by2 \advance \xpos by\coefa \advance \xpos by\width
\multiply \coefa by\rise \divide \coefa by\run \advance \ypos
by\coefa
\if l#9%
   \putrbox(\xpos,\ypos){#6}%
\else\if r#9%
   \putlbox(\xpos,\ypos){#6}%
\fi\fi }%
{\multiply \rise by-\coefc
\multiply \run by-\coefc \multiply \coefb by\arrowlength \advance
\xpos by\coefb \multiply \coefb by\rise \divide \coefb by\run
\advance \ypos by\coefb \multiply \coefc by70 \advance \ypos
by\coefc \multiply \coefc by\run \divide \coefc by\rise \advance
\xpos by\coefc \multiply \coefa by140 \multiply \coefa by\run
\divide \coefa by\rise \advance \arrowlength by\coefa
\ifcase\arrowtype
\or \put(\xpos,\ypos){\vector(\run,\rise){\arrowlength}}%
\or \put(\xpos,\ypos){\mvector(\run,\rise){\arrowlength}}%
\or \put(\xpos,\ypos){\evector(\run,\rise){\arrowlength}}%
\fi}\fi\fi\fi\fi}}

\newcount\numbdashes \newcount\lengthdash \newcount\increment

\def\howmanydashes{
\numbdashes=\arrowlength \lengthdash=40 \divide\numbdashes by
\lengthdash \lengthdash=\arrowlength \divide\lengthdash by
\numbdashes
\increment=\lengthdash \multiply\lengthdash by 3
\divide\lengthdash by 5 }

\def\putdashvector(#1)(#2,#3)#4#5{%
\ifnum#3=0 \putdashhvector(#1){#4}#5 \else \ifnum#2=0
\putdashvvector(#1){#4}#5\fi\fi}

\def\putdashhvector(#1,#2)#3#4{{%
\arrowlength=#3 \howmanydashes
\multiput(#1,#2)(\increment,0){\numbdashes}%
{\vrule height .4pt width \lengthdash\unitlength} \arrowtype=#4
\xpos=#1 \ifnum\arrowtype<0 \advance\arrowtype by 7 \fi
\ifcase\arrowtype \or \advance\xpos by 10
    \put(\xpos,#2){\vector(-1,0){\lengthdash}}
    \advance\xpos by 40
    \put(\xpos,#2){\vector(-1,0){\lengthdash}}
\or \advance \xpos by 10
    \put(\xpos,#2){\vector(-1,0){\lengthdash}}
    \advance\xpos by  \arrowlength
    \advance\xpos by  -50
    \put(\xpos,#2){\vector(-1,0){\lengthdash}}
\or \advance\xpos by 10
    \put(\xpos,#2){\vector(-1,0){\lengthdash}}
\or \advance\xpos by \arrowlength
    \advance\xpos by -\lengthdash
    \put(\xpos,#2){\vector(1,0){\lengthdash}}
\or {\advance\xpos by 10
    \put(\xpos,#2){\vector(1,0){\lengthdash}}}
    \advance\xpos by \arrowlength
    \advance\xpos by -\lengthdash
    \put(\xpos,#2){\vector(1,0){\lengthdash}}
\or \advance\xpos by \arrowlength
    \advance\xpos by -\lengthdash
    \put(\xpos,#2){\vector(1,0){\lengthdash}}
    \advance\xpos by -40
    \put(\xpos,#2){\vector(1,0){\lengthdash}}
   \fi
}}

\def\putdashvvector(#1,#2)#3#4{{%
\arrowlength=#3 \howmanydashes \ypos=#2 \advance\ypos by
-\arrowlength
\multiput(#1,#2)(0,\increment){\numbdashes}%
    {\vrule width .4pt height \lengthdash\unitlength}
\arrowtype=#4 \ypos=#2 \ifnum\arrowtype<0 \advance\arrowtype by 7
\fi \ifcase\arrowtype \or \advance\ypos by \arrowlength
\advance\ypos by -40
    \put(#1,\ypos){\vector(0,1){\lengthdash}}
    \advance\ypos by -40
    \put(#1,\ypos){\vector(0,1){\lengthdash}}
\or \advance\ypos by 10
    \put(#1,\ypos){\vector(0,1){\lengthdash}}
    \advance\ypos by \arrowlength \advance\ypos by -40
    \put(#1,\ypos){\vector(0,1){\lengthdash}}
\or \advance\ypos by \arrowlength \advance\ypos by -40
    \put(#1,\ypos){\vector(0,1){\lengthdash}}
\or \advance\ypos by 10
    \put(#1,\ypos){\vector(0,-1){\lengthdash}}
\or \advance\ypos by 10
    \put(#1,\ypos){\vector(0,-1){\lengthdash}}
    \advance\ypos by \arrowlength \advance\ypos by -40
    \put(#1,\ypos){\vector(0,-1){\lengthdash}}
\or \advance\ypos by 10
    \put(#1,\ypos){\vector(0,-1){\lengthdash}}
    \advance\ypos by 40
    \put(#1,\ypos){\vector(0,-1){\lengthdash}}
\fi }}

\def\puthmorphism(#1,#2)[#3`#4`#5]#6#7#8{{%
\xpos #1 \ypos #2 \width #6 \arrowlength #6 \arrowtype=#7
\putbox(\xpos,\ypos){#3\vphantom{#4}}%
{\advance \xpos by\arrowlength
\putbox(\xpos,\ypos){\vphantom{#3}#4}}%
\horsize{\tempcounta}{#3}%
\horsize{\tempcountb}{#4}%
\divide \tempcounta by2 \divide \tempcountb by2 \advance
\tempcounta by30 \advance \tempcountb by30 \advance \xpos
by\tempcounta \advance \arrowlength by-\tempcounta \advance
\arrowlength by-\tempcountb
\putvector(\xpos,\ypos)(1,0)\arrowlength\arrowtype \divide
\arrowlength by2 \advance \xpos by\arrowlength
\vertsize{\tempcounta}{#5}%
\divide\tempcounta by2 \advance \tempcounta by20
\if a#8 %
   \advance \ypos by\tempcounta
   \putbox(\xpos,\ypos){#5}%
\else
   \advance \ypos by-\tempcounta
   \putbox(\xpos,\ypos){#5}%
\fi}}

\def\putvmorphism(#1,#2)[#3`#4`#5]#6#7#8{{%
\xpos #1 \ypos #2 \arrowlength #6 \arrowtype #7
\settowidth{\xlen}{$#5$}%
\putbox(\xpos,\ypos){#3}%
{\advance \ypos by-\arrowlength
\putbox(\xpos,\ypos){#4}}%
{\advance\arrowlength by-140 \advance \ypos by-70 \ifdim\xlen>0pt
   \if m#8%
      \putsplitvector(\xpos,\ypos)\arrowlength\arrowtype
   \else
   \putvector(\xpos,\ypos)(0,-1)\arrowlength\arrowtype
   \fi
\else
   \putvector(\xpos,\ypos)(0,-1)\arrowlength\arrowtype
\fi}%
\ifdim\xlen>0pt
   \divide \arrowlength by2
   \advance\ypos by-\arrowlength
   \if l#8%
      \advance \xpos by-40
      \putrbox(\xpos,\ypos){#5}%
   \else\if r#8%
      \advance \xpos by40
      \putlbox(\xpos,\ypos){#5}%
   \else
      \putbox(\xpos,\ypos){#5}%
   \fi\fi
\fi }}

\def\putsquarep<#1>(#2)[#3;#4`#5`#6`#7]{{%
\setsqparms[#1]%
\setpos(#2)%
\settokens`#3`%
\puthmorphism(\xpos,\ypos)[\tokenc`\tokend`{#7}]{\width}{\arrowtyped}b%
\advance\ypos by \height
\puthmorphism(\xpos,\ypos)[\tokena`\tokenb`{#4}]{\width}{\arrowtypea}a%
\putvmorphism(\xpos,\ypos)[``{#5}]{\height}{\arrowtypeb}l%
\advance\xpos by \width
\putvmorphism(\xpos,\ypos)[``{#6}]{\height}{\arrowtypec}r%
}}

\def\putsquare{\@ifnextchar <{\putsquarep}{\putsquarep%
   <\arrowtypea`\arrowtypeb`\arrowtypec`\arrowtyped;\width`\height>}}
\def\square{\@ifnextchar< {\squarep}{\squarep
   <\arrowtypea`\arrowtypeb`\arrowtypec`\arrowtyped;\width`\height>}}
\def\squarep<#1>[#2`#3`#4`#5;#6`#7`#8`#9]{{
\setsqparms[#1]
\diagram
\putsquarep<\arrowtypea`\arrowtypeb`\arrowtypec`
\arrowtyped;\width`\height>
(0,0)[#2`#3`#4`{#5};#6`#7`#8`{#9}]
\enddiagram
}}                                                 
\def\putptrianglep<#1>(#2,#3)[#4`#5`#6;#7`#8`#9]{{%
\settriparms[#1]%
\xpos=#2 \ypos=#3 \advance\ypos by \height
\puthmorphism(\xpos,\ypos)[#4`#5`{#7}]{\height}{\arrowtypea}a%
\putvmorphism(\xpos,\ypos)[`#6`{#8}]{\height}{\arrowtypeb}l%
\advance\xpos by\height
\putmorphism(\xpos,\ypos)(-1,-1)[``{#9}]{\height}{\arrowtypec}r%
}}

\def\putptriangle{\@ifnextchar <{\putptrianglep}{\putptrianglep
   <\arrowtypea`\arrowtypeb`\arrowtypec;\height>}}
\def\ptriangle{\@ifnextchar <{\ptrianglep}{\ptrianglep
   <\arrowtypea`\arrowtypeb`\arrowtypec;\height>}}
\def\ptrianglep<#1>[#2`#3`#4;#5`#6`#7]{{
\settriparms[#1]
\diagram
\putptrianglep<\arrowtypea`\arrowtypeb`
\arrowtypec;\height>
(0,0)[#2`#3`#4;#5`#6`{#7}]
\enddiagram
}}                                            

\def\putqtrianglep<#1>(#2,#3)[#4`#5`#6;#7`#8`#9]{{%
\settriparms[#1]%
\xpos=#2 \ypos=#3 \advance\ypos by\height
\puthmorphism(\xpos,\ypos)[#4`#5`{#7}]{\height}{\arrowtypea}a%
\putmorphism(\xpos,\ypos)(1,-1)[``{#8}]{\height}{\arrowtypeb}l%
\advance\xpos by\height
\putvmorphism(\xpos,\ypos)[`#6`{#9}]{\height}{\arrowtypec}r%
}}

\def\putqtriangle{\@ifnextchar <{\putqtrianglep}{\putqtrianglep
   <\arrowtypea`\arrowtypeb`\arrowtypec;\height>}}
\def\qtriangle{\@ifnextchar <{\qtrianglep}{\qtrianglep
   <\arrowtypea`\arrowtypeb`\arrowtypec;\height>}}
\def\qtrianglep<#1>[#2`#3`#4;#5`#6`#7]{{
\settriparms[#1]
\width=\height                                
\diagram
\putqtrianglep<\arrowtypea`\arrowtypeb`
\arrowtypec;\height>
(0,0)[#2`#3`#4;#5`#6`{#7}]
\enddiagram
}}

\def\putdtrianglep<#1>(#2,#3)[#4`#5`#6;#7`#8`#9]{{%
\settriparms[#1]%
\xpos=#2 \ypos=#3
\puthmorphism(\xpos,\ypos)[#5`#6`{#9}]{\height}{\arrowtypec}b%
\advance\xpos by \height \advance\ypos by\height
\putmorphism(\xpos,\ypos)(-1,-1)[``{#7}]{\height}{\arrowtypea}l%
\putvmorphism(\xpos,\ypos)[#4``{#8}]{\height}{\arrowtypeb}r%
}}

\def\putdtriangle{\@ifnextchar <{\putdtrianglep}{\putdtrianglep
   <\arrowtypea`\arrowtypeb`\arrowtypec;\height>}}
\def\dtriangle{\@ifnextchar <{\dtrianglep}{\dtrianglep
   <\arrowtypea`\arrowtypeb`\arrowtypec;\height>}}
\def\dtrianglep<#1>[#2`#3`#4;#5`#6`#7]{{
\settriparms[#1]
\width=\height                                
\diagram
\putdtrianglep<\arrowtypea`\arrowtypeb`
\arrowtypec;\height>
(0,0)[#2`#3`#4;#5`#6`{#7}]
\enddiagram
}}

\def\putbtrianglep<#1>(#2,#3)[#4`#5`#6;#7`#8`#9]{{%
\settriparms[#1]%
\xpos=#2 \ypos=#3
\puthmorphism(\xpos,\ypos)[#5`#6`{#9}]{\height}{\arrowtypec}b%
\advance\ypos by\height
\putmorphism(\xpos,\ypos)(1,-1)[``{#8}]{\height}{\arrowtypeb}r%
\putvmorphism(\xpos,\ypos)[#4``{#7}]{\height}{\arrowtypea}l%
}}

\def\putbtriangle{\@ifnextchar <{\putbtrianglep}{\putbtrianglep
   <\arrowtypea`\arrowtypeb`\arrowtypec;\height>}}
\def\btriangle{\@ifnextchar <{\btrianglep}{\btrianglep
   <\arrowtypea`\arrowtypeb`\arrowtypec;\height>}}
\def\btrianglep<#1>[#2`#3`#4;#5`#6`#7]{{
\settriparms[#1]
\width=\height                               
\diagram
\putbtrianglep<\arrowtypea`\arrowtypeb`
\arrowtypec;\height>
(0,0)[#2`#3`#4;#5`#6`{#7}]
\enddiagram
}}

\def\putAtrianglep<#1>(#2,#3)[#4`#5`#6;#7`#8`#9]{{%
\settriparms[#1]%
\xpos=#2 \ypos=#3 {\multiply \height by2
\puthmorphism(\xpos,\ypos)[#5`#6`{#9}]{\height}{\arrowtypec}b}%
\advance\xpos by\height \advance\ypos by\height
\putmorphism(\xpos,\ypos)(-1,-1)[#4``{#7}]{\height}{\arrowtypea}l%
\putmorphism(\xpos,\ypos)(1,-1)[``{#8}]{\height}{\arrowtypeb}r%
}}

\def\putAtriangle{\@ifnextchar <{\putAtrianglep}{\putAtrianglep
   <\arrowtypea`\arrowtypeb`\arrowtypec;\height>}}
\def\Atriangle{\@ifnextchar <{\Atrianglep}{\Atrianglep
   <\arrowtypea`\arrowtypeb`\arrowtypec;\height>}}
\def\Atrianglep<#1>[#2`#3`#4;#5`#6`#7]{{
\settriparms[#1]
\width=\height                                     
\diagram
\putAtrianglep<\arrowtypea`\arrowtypeb`
\arrowtypec;\height>
(0,0)[#2`#3`#4;#5`#6`{#7}]
\enddiagram
}}

\def\putAtrianglepairp<#1>(#2)[#3;#4`#5`#6`#7`#8]{{%
\settripairparms[#1]%
\setpos(#2)%
\settokens`#3`%
\puthmorphism(\xpos,\ypos)[\tokenb`\tokenc`{#7}]{\height}{\arrowtyped}b%
\advance\xpos by\height
\puthmorphism(\xpos,\ypos)[\phantom{\tokenc}`\tokend`{#8}]%
{\height}{\arrowtypee}b%
\advance\ypos by\height
\putmorphism(\xpos,\ypos)(-1,-1)[\tokena``{#4}]{\height}{\arrowtypea}l%
\putvmorphism(\xpos,\ypos)[``{#5}]{\height}{\arrowtypeb}m%
\putmorphism(\xpos,\ypos)(1,-1)[``{#6}]{\height}{\arrowtypec}r%
}}

\def\putAtrianglepair{\@ifnextchar <{\putAtrianglepairp}{\putAtrianglepairp%
   <\arrowtypea`\arrowtypeb`\arrowtypec`\arrowtyped`\arrowtypee;\height>}}
\def\Atrianglepair{\@ifnextchar <{\Atrianglepairp}{\Atrianglepairp%
   <\arrowtypea`\arrowtypeb`\arrowtypec`\arrowtyped`\arrowtypee;\height>}}

\def\Atrianglepairp<#1>[#2;#3`#4`#5`#6`#7]{{
\settripairparms[#1]
\settokens`#2`
\width=\height                                
\diagram
\putAtrianglepairp                            
<\arrowtypea`\arrowtypeb`\arrowtypec`
\arrowtyped`\arrowtypee;\height>
(0,0)[{#2};#3`#4`#5`#6`{#7}]
\enddiagram
}}

\def\putVtrianglep<#1>(#2,#3)[#4`#5`#6;#7`#8`#9]{{%
\settriparms[#1]%
\xpos=#2 \ypos=#3 \advance\ypos by\height {\multiply\height by2
\puthmorphism(\xpos,\ypos)[#4`#5`{#7}]{\height}{\arrowtypea}a}%
\putmorphism(\xpos,\ypos)(1,-1)[`#6`{#8}]{\height}{\arrowtypeb}l%
\advance\xpos by\height \advance\xpos by\height
\putmorphism(\xpos,\ypos)(-1,-1)[``{#9}]{\height}{\arrowtypec}r%
}}

\def\putVtriangle{\@ifnextchar <{\putVtrianglep}{\putVtrianglep
   <\arrowtypea`\arrowtypeb`\arrowtypec;\height>}}
\def\Vtriangle{\@ifnextchar <{\Vtrianglep}{\Vtrianglep
   <\arrowtypea`\arrowtypeb`\arrowtypec;\height>}}
\def\Vtrianglep<#1>[#2`#3`#4;#5`#6`#7]{{
\settriparms[#1]
\width=\height                                 
\diagram
\putVtrianglep<\arrowtypea`\arrowtypeb`
\arrowtypec;\height>
(0,0)[#2`#3`#4;#5`#6`{#7}]
\enddiagram
}}

\def\putVtrianglepairp<#1>(#2)[#3;#4`#5`#6`#7`#8]{{
\settripairparms[#1]%
\setpos(#2)%
\settokens`#3`%
\advance\ypos by\height
\putmorphism(\xpos,\ypos)(1,-1)[`\tokend`{#6}]{\height}{\arrowtypec}l%
\puthmorphism(\xpos,\ypos)[\tokena`\tokenb`{#4}]{\height}{\arrowtypea}a%
\advance\xpos by\height
\puthmorphism(\xpos,\ypos)[\phantom{\tokenb}`\tokenc`{#5}]%
{\height}{\arrowtypeb}a%
\putvmorphism(\xpos,\ypos)[``{#7}]{\height}{\arrowtyped}m%
\advance\xpos by\height
\putmorphism(\xpos,\ypos)(-1,-1)[``{#8}]{\height}{\arrowtypee}r%
}}

\def\putVtrianglepair{\@ifnextchar <{\putVtrianglepairp}{\putVtrianglepairp%
    <\arrowtypea`\arrowtypeb`\arrowtypec`\arrowtyped`\arrowtypee;\height>}}
\def\Vtrianglepair{\@ifnextchar <{\Vtrianglepairp}{\Vtrianglepairp%
    <\arrowtypea`\arrowtypeb`\arrowtypec`\arrowtyped`\arrowtypee;\height>}}
\def\Vtrianglepairp<#1>[#2;#3`#4`#5`#6`#7]{{
\settripairparms[#1]
\settokens`#2`
\diagram
\putVtrianglepairp                             
<\arrowtypea`\arrowtypeb`\arrowtypec`
\arrowtyped`\arrowtypee;\height>
(0,0)[{#2};#3`#4`#5`#6`{#7}]
\enddiagram
}}

\def\putCtrianglep<#1>(#2,#3)[#4`#5`#6;#7`#8`#9]{{%
\settriparms[#1]%
\xpos=#2 \ypos=#3 \advance\ypos by\height
\putmorphism(\xpos,\ypos)(1,-1)[``{#9}]{\height}{\arrowtypec}l%
\advance\xpos by\height \advance\ypos by\height
\putmorphism(\xpos,\ypos)(-1,-1)[#4`#5`{#7}]{\height}{\arrowtypea}l%
{\multiply\height by 2
\putvmorphism(\xpos,\ypos)[`#6`{#8}]{\height}{\arrowtypeb}r}%
}}

\def\putCtriangle{\@ifnextchar <{\putCtrianglep}{\putCtrianglep
    <\arrowtypea`\arrowtypeb`\arrowtypec;\height>}}
\def\Ctriangle{\@ifnextchar <{\Ctrianglep}{\Ctrianglep
    <\arrowtypea`\arrowtypeb`\arrowtypec;\height>}}
\def\Ctrianglep<#1>[#2`#3`#4;#5`#6`#7]{{
\settriparms[#1]
\width=\height                               
\diagram
\putCtrianglep<\arrowtypea`\arrowtypeb`
\arrowtypec;\height>
(0,0)[#2`#3`#4;#5`#6`{#7}]
\enddiagram
}}                                           
\def\putDtrianglep<#1>(#2,#3)[#4`#5`#6;#7`#8`#9]{{%
\settriparms[#1]%
\xpos=#2 \ypos=#3 \advance\xpos by\height \advance\ypos by\height
\putmorphism(\xpos,\ypos)(-1,-1)[``{#9}]{\height}{\arrowtypec}r%
\advance\xpos by-\height \advance\ypos by\height
\putmorphism(\xpos,\ypos)(1,-1)[`#5`{#8}]{\height}{\arrowtypeb}r%
{\multiply\height by 2
\putvmorphism(\xpos,\ypos)[#4`#6`{#7}]{\height}{\arrowtypea}l}%
}}

\def\putDtriangle{\@ifnextchar <{\putDtrianglep}{\putDtrianglep
    <\arrowtypea`\arrowtypeb`\arrowtypec;\height>}}
\def\Dtriangle{\@ifnextchar <{\Dtrianglep}{\Dtrianglep
   <\arrowtypea`\arrowtypeb`\arrowtypec;\height>}}
\def\Dtrianglep<#1>[#2`#3`#4;#5`#6`#7]{{
\settriparms[#1]
\width=\height                              
\diagram
\putDtrianglep<\arrowtypea`\arrowtypeb`
\arrowtypec;\height>
(0,0)[#2`#3`#4;#5`#6`{#7}]
\enddiagram
}}                                          
\def\setrecparms[#1`#2]{\width=#1 \height=#2}%

\def\recursep<#1`#2>[#3;#4`#5`#6`#7`#8]{{\m@th
\width=#1 \height=#2 \settokens`#3`
\settowidth{\tempdimen}{$\tokena$} \ifdim\tempdimen=0pt
  \savebox{\tempboxa}{\hbox{$\tokenb$}}%
  \savebox{\tempboxb}{\hbox{$\tokend$}}%
  \savebox{\tempboxc}{\hbox{$#6$}}%
\else
  \savebox{\tempboxa}{\hbox{$\hbox{$\tokena$}\times\hbox{$\tokenb$}$}}%
  \savebox{\tempboxb}{\hbox{$\hbox{$\tokena$}\times\hbox{$\tokend$}$}}%
  \savebox{\tempboxc}{\hbox{$\hbox{$\tokena$}\times\hbox{$#6$}$}}%
\fi \ypos=\height \divide\ypos by 2 \xpos=\ypos \advance\xpos by
\width \bfig
\putCtrianglep<-1`1`1;\ypos>(0,0)[`\tokenc`;#5`#6`{#7}]%
\puthmorphism(\ypos,0)[\tokend`\usebox{\tempboxb}`{#8}]{\width}{-1}b%
\puthmorphism(\ypos,\height)[\tokenb`\usebox{\tempboxa}`{#4}]{\width}{-1}a%
\advance\ypos by \width
\putvmorphism(\ypos,\height)[``\usebox{\tempboxc}]{\height}1r%
\efig }}

\def\recurse{\@ifnextchar <{\recursep}{\recursep<\width`\height>}}

\def\puttwohmorphisms(#1,#2)[#3`#4;#5`#6]#7#8#9{{%
\puthmorphism(#1,#2)[#3`#4`]{#7}0a \ypos=#2 \advance\ypos by 20
\puthmorphism(#1,\ypos)[\phantom{#3}`\phantom{#4}`#5]{#7}{#8}a
\advance\ypos by -40
\puthmorphism(#1,\ypos)[\phantom{#3}`\phantom{#4}`#6]{#7}{#9}b }}

\def\puttwovmorphisms(#1,#2)[#3`#4;#5`#6]#7#8#9{{%
\putvmorphism(#1,#2)[#3`#4`]{#7}0a \xpos=#1 \advance\xpos by -20
\putvmorphism(\xpos,#2)[\phantom{#3}`\phantom{#4}`#5]{#7}{#8}l
\advance\xpos by 40
\putvmorphism(\xpos,#2)[\phantom{#3}`\phantom{#4}`#6]{#7}{#9}r }}

\def\puthcoequalizer(#1)[#2`#3`#4;#5`#6`#7]#8#9{{%
\setpos(#1)%
\puttwohmorphisms(\xpos,\ypos)[#2`#3;#5`#6]{#8}11%
\advance\xpos by #8
\puthmorphism(\xpos,\ypos)[\phantom{#3}`#4`#7]{#8}1{#9} }}

\def\putvcoequalizer(#1)[#2`#3`#4;#5`#6`#7]#8#9{{%
\setpos(#1)%
\puttwovmorphisms(\xpos,\ypos)[#2`#3;#5`#6]{#8}11%
\advance\ypos by -#8
\putvmorphism(\xpos,\ypos)[\phantom{#3}`#4`#7]{#8}1{#9} }}

\def\putthreehmorphisms(#1)[#2`#3;#4`#5`#6]#7(#8)#9{{%
\setpos(#1) \settypes(#8)
\if a#9 %
     \vertsize{\tempcounta}{#5}%
     \vertsize{\tempcountb}{#6}%
     \ifnum \tempcounta<\tempcountb \tempcounta=\tempcountb \fi
\else
     \vertsize{\tempcounta}{#4}%
     \vertsize{\tempcountb}{#5}%
     \ifnum \tempcounta<\tempcountb \tempcounta=\tempcountb \fi
\fi \advance \tempcounta by 60
\puthmorphism(\xpos,\ypos)[#2`#3`#5]{#7}{\arrowtypeb}{#9}
\advance\ypos by \tempcounta
\puthmorphism(\xpos,\ypos)[\phantom{#2}`\phantom{#3}`#4]{#7}{\arrowtypea}{#9}
\advance\ypos by -\tempcounta \advance\ypos by -\tempcounta
\puthmorphism(\xpos,\ypos)[\phantom{#2}`\phantom{#3}`#6]{#7}{\arrowtypec}{#9}
}}

\def\setarrowtoks[#1`#2`#3`#4`#5`#6]{%
\def\toka{#1}
\def\tokb{#2}
\def\tokc{#3}
\def\tokd{#4}
\def\toke{#5}
\def\tokf{#6}
}
\def\hex{\@ifnextchar <{\hexp}{\hexp<1000`400>}}
\def\hexp<#1`#2>[#3`#4`#5`#6`#7`#8;#9]{%
\setarrowtoks[#9] \yext=#2 \advance \yext by #2 \xext=#1
\advance\xext by \yext \bfig
\putCtriangle<-1`0`1;#2>(0,0)[`#5`;\tokb``\tokd] \xext=#1 \yext=#2
\advance \yext by #2
\putsquare<1`0`0`1;\xext`\yext>(#2,0)[#3`#4`#7`#8;\toka```\tokf]
\advance \xext by #2
\putDtriangle<0`1`-1;#2>(\xext,0)[`#6`;`\tokc`\toke] \efig }
\makeatother